\documentclass[reprint,amsmath,amssymb,aps]{revtex4-2}
\usepackage{graphicx}
\usepackage{dcolumn}
\usepackage{bm}

\begin{document}
	\preprint{APS/123-QED}
\title{Granular packings with sliding, rolling and twisting friction}
	
\author{A. P. Santos}
\email{asanto@sandia.gov}
\author{Dan S. Bolintineanu}
\author{Gary S. Grest}
\author{Jeremy B. Lechman}
\author{Steven J. Plimpton}
\affiliation{Sandia National Laboratories, Albuquerque, NM 87185, USA}
\author{Leonardo E. Silbert}
\affiliation{School of Math, Science and Engineering, Central New Mexico Community College, Albuquerque, NM 87106, USA}
\author{Ishan Srivastava} \altaffiliation[Present address: ]{Center for Computational Sciences and Engineering, Lawrence Berkeley National Laboratory, Berkeley, California 94720, USA}

\affiliation{Sandia National Laboratories, Albuquerque, NM 87185, USA}
\begin{abstract}
Intuition tells us that a rolling or spinning sphere will eventually stop due to the presence of friction and other dissipative interactions. The resistance to rolling and spinning/twisting torque that stops a sphere also changes the microstructure of a granular packing of frictional spheres by increasing the number of constraints on the degrees of freedom of motion.  We perform discrete element modeling simulations to construct sphere packings implementing a range of frictional constraints under a pressure-controlled protocol. Mechanically stable packings are achievable at volume fractions and average coordination numbers as low as 0.53 and 2.5, respectively, when the particles experience high resistance to sliding, rolling and twisting. Only when the particle model includes rolling and twisting friction, were experimental volume fractions reproduced.
\end{abstract}
\maketitle
\section{Introduction}\label{sec:intro}
A rolling or spinning marble on a table eventually slows to a stop because of resistance to the rolling and twisting modes of motion. However, rolling and twisting friction are often excluded in simulation studies because of the added complexity of the contact mechanics model.  Such approximations may be valid for some phenomena and materials, such as materials with low sliding friction~\cite{Skinner1969}, but the validity of this approximation, and the ability to match experimental properties,  must be tested. 


Simulations have found that rolling and twisting friction is necessary to reproduce experimental observations and can change macroscopic behavior. Only by including rolling friction were Mort et al.~\cite{Mort2015} able to reproduce experimental shear/normal stress ratios in a hopper. Singh et al.~\cite{Singh2020} were also unable to reproduce experimental shear viscosity at relevant sliding friction coefficients with simulations without rolling friction. 
Other simulations found that rolling friction induces columnar granular particle contact backbones~\cite{Estrada2008}, increases stress-dilatancy~\cite{Liu2018e}, causes anisotropic dense granular flows~\cite{Wu2019} and provides an explanation of discontinuous shear thickening~\cite{Mari2019,Guy2018}.
Simulations of shear banding~\cite{Bardet1994,Iwashita1998}, rigid flat-punch~\cite{Tordesillas2016} and wing-crack extension~\cite{Wang2008a} processes generated large regions of rotations. Resistance to such rotation could change  behavior, and studies of such processes should consider including rolling and twisting friction in models. 
The magnitude of resistance to rolling and twisting can be approximated, and explained by, contact mechanics theory.

Long before granular particles were simulated with rolling and twisting friction, Reynolds~\cite{Reynolds1875} and Hertz~\cite{Hertz1882} used theories of rolling and twisting resistance to suggest a substantial impact on packing structure. These theories focus on single particle-wall interactions using elastic and inelastic approximations of rolling resistance~\cite{Brilliantov1998,Poschel1999}. 
Twisting and sliding friction have the same origins -- twisting having rotational instead of translational displacements over the contact area.  Rolling friction originates from a combination of micro-slip at the interface, inelastic deformation and surface roughness that create a pressure difference between the leading and trailing ends of the rolling contact~\cite{Johnson1985}. 
Constraint counting predicts that rolling and twisting resistance leads to large changes in packing structure (see Sec.~\ref{sec:constraint_counting}).
Particle properties, including surface morphology and the material~\cite{Johnson1985}, sets the rolling and twisting resistance, and thus can control packing structure. 

Packings of spheres and disks with sliding friction have shown many interesting phenomena. 
For example, in 3 dimensions, the coordination number $Z$ decreases gradually with friction from the frictionless value of $Z$=6 to the frictional isostatic number $Z$=4~\cite{Silbert2002,Shundyak2007,Somfai2007,Song2008a,Silbert2010}.
The volume fraction $\phi$ follows the decrease in coordination number with increasing friction.
Frictionless hard spheres sets the densest volume fraction $\phi\!\sim\!0.64$ for random packings, known as the maximally random jammed state~\cite{Torquato2000}.
Simulations~\cite{Silbert2002} and theory~\cite{Song2008a} have shown that sliding friction leads to looser granular sphere packings than frictionless sphere packings, as verified by experiments~\cite{Scott1969}. 

How loose frictional packings can be depends on the friction coefficient and the path to packing. Early packing experiments of monodisperse spheres measured a range of volume fractions with minimums of $\phi\sim$0.57-0.6~\cite{Brown1946,Scott1960,Rutgers1962}. Later experiments demonstrated that materials with larger  friction coefficients can access mechanically stable packings with lower volume fractions~\cite{Scott1969,Jerkins2008,Farrell2010}.  With careful, density matched experiments, Farrell et al.~\cite{Farrell2010} measured volume fractions $\phi<$0.54 for very frictional particles.
Silbert~\cite{Silbert2010} used a volume-controlled simulation protocol to produce stable packings as low as $\phi=$0.576 for high, but realistic sliding friction coefficient values ($\mu_s=0.5$), which is still above the experimental value. Because rolling and twisting friction makes models more realistic, it offers simulations a route to match the low volume fractions seen in experimental packings.

Packings generated with a volume-controlled protocol have difficulty forming stable packings at low pressures and volume fractions. By controlling pressure not volume, final packings repeatably and rigorously satisfy set stress conditions at low pressures. Pressure-controlled protocols have been used to study granular packings~\cite{Shundyak2007,Dagois-Bohy2012,Smith2014,Srivastava2017}, but their application to 3D frictional particles is less common. In this article, a constant pressure in the x-, y- and z-directions allows the box to adjust the edge length, and by allowing the box to adopt triclinic configurations, constant zero shear stresses is achieved. Packings formed by pressure-controlled protocols are more stable to shear deformation than volume-controlled methods, as shown by Dagois-Bohy et al.~\cite{Dagois-Bohy2012} and Smith et al.~\cite{Smith2014}.

In this article, a constant-pressure protocol is used to pack three-dimensional, monodisperse particles with varying degrees of sliding, rolling and twisting friction. 
The equations of motion that define contact forces for the normal, sliding, rolling and twisting modes are presented in Sec.~\ref{sec:model}. The appropriate magnitudes of the rolling and twisting contact force parameters are discussed in Appendix~\ref{sec:appendix}. The details and benefits of the constant-pressure packing protocol are described in Sec.~\ref{sec:boxeom}. The effect of rolling and twisting friction on packing morphology is first predicted using constraint counting (Sec.~\ref{sec:constraint_counting}), and then the results of numerical simulations are presented and compared to experiment in Sec.~\ref{sec:frictions}.

\section{Methodology}\label{sec:methods} 
\subsection{Contact model}\label{sec:model} 
Granular particles are modeled as spherical particles with radius $R_i$ and mass $m_i$.  Particles only interact when in contact, through a spring-dashpot-slider interaction potential for the normal, sliding, rolling and twisting modes of motion. Sliding friction uses the Cundall and Strack model~\cite{Cundall1979}.  Rolling resistance is based on Luding's implementation~\cite{Luding2008} and twisting resistence is based on Marshall's implementation~\cite{Marshall2009}.  For two granular particles in contact, separated by a distance $|\mathbf{r}_{ij}| < R_i+R_j$, the force on particle $i$ from particle $j$ is
\begin{subequations}
\begin{align}
\mathbf{F}_{ij} =& \mathbf{F}_n + \mathbf{F}_s \label{eqn:Ftotal}\\
\mathbf{F}_n =& k_n \delta \mathbf{n} - m_{\text{eff}}\gamma_n \mathbf{v}_n \\
\mathbf{F}_s =& -\text{min}\left( \mu_s |\mathbf{F}_n|, | -k_s \boldsymbol\xi_{s} - m_{\text{eff}}\gamma_s \mathbf{v}_s |\right)\frac{\mathbf{v}_s}{|\mathbf{v}_s|}  \label{eqn:Fs}
\end{align}
\end{subequations}
\noindent where $\mathbf{\delta} = R_i + R_j - |\mathbf{r}_{ij}|$, $\mathbf{n} = \frac{\mathbf{r}_{ij}}{|\mathbf{r}_{ij}|}$, $m_{\text{eff}} = \frac{m_i m_j}{m_i + m_j}$. $\mathbf{F}_n$ is the normal force and $\mathbf{F}_s$ is the sliding force. 
Sliding ($s$), rolling ($r$) and twisting ($t$) give rise to torque when granular particles are in contact. The torque acting on particle $i$ due to contact with particle $j$ is defined as:
\begin{subequations}
\begin{align}
\boldsymbol{\tau}_{ij} =& \boldsymbol{\tau}_s +  \boldsymbol{\tau}_r + \boldsymbol\tau_t \label{eqn:Ttotal}\\
\boldsymbol{\tau}_s =& -\left(R_i-\frac{\delta}{2}\right)\mathbf{n} \times \mathbf{F}_s \\ 
\boldsymbol{\tau}_r =& -R_{\text{eff}}\mathbf{n} \times \text{min}\left( \mu_r |\mathbf{F}_n|, | -k_r \boldsymbol\xi_{r} - \gamma_r \mathbf{v}_r |\right)\frac{\mathbf{v}_r}{|\mathbf{v}_r|}  \\ 
\boldsymbol\tau_t =& -\text{min}\left( \mu_t |\mathbf{F}_n|, -k_t \xi_{t} - \gamma_t v_t\right) \mathbf{n}  \label{eqn:Ft}
\end{align}
\end{subequations}
\noindent where $R_{\text{eff}}=\frac{R_iR_j}{R_i+R_j}$. Note that torque acting on particle $j$ due to contact with particle $i$ is $\boldsymbol{\tau}_{ji}=-\boldsymbol{\tau}_{ij}$, except for $\boldsymbol{\tau}_s$ if $R_i\ne R_j$. Each mode of motion $m$ has a Hookean spring constant $k_m$ and viscoelastic damping coefficient $\gamma_m$, that takes into account the inelasticity of the contact mechanics.  The Coulomb yield criteria is applied to each frictional mode force or torque and sets the maximum to be the friction coefficient $\mu_m$ times the normal force. 

\begin{figure}[!htbp]
	\centering
	\includegraphics[width=\columnwidth]{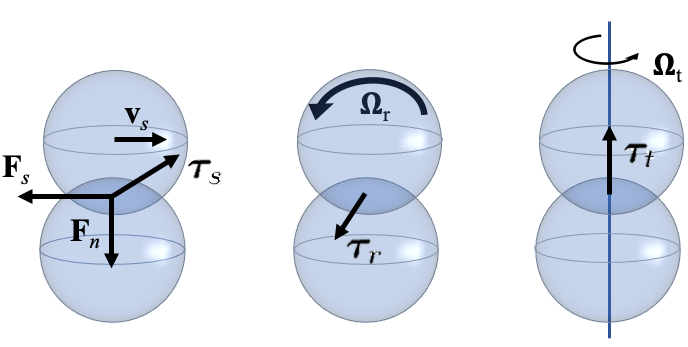}
	\caption{Schematic of the granular particle interaction model normal $\mathbf{F}_n$ and sliding $\mathbf{F}_s$ forces, and the sliding $\boldsymbol\tau_s$, rolling $\boldsymbol\tau_r$ and twisting $\boldsymbol\tau_t$ torques. Resistence to the sliding translational velocity $\mathbf{v}_s$,  the rolling $\mathbf{\Omega}_r$ and twisting $\mathbf{\Omega}_t$ rotational velocities cause frictional forces and torques.}
	\label{fig:Modelfig}
\end{figure}

The displacement accumulated as particles are in contact is an important aspect of this model because it captures micro-slip and has been observed in experimental studies of oblique impact~\cite{Foerster1994}. 
The accumulated displacement is measured by $\boldsymbol{\xi}_m = \int_{t_0}^t \mathbf{v}_{m}(\tau) \mathrm{d}\tau$, where $t_0$ is the time at first contact. 
To compensate for the effect of rigid body rotations, $\boldsymbol{\xi}_m$ is calculated in the reference frame of the rotating particle pair~\cite{Luding2008}.  As the contacting pair rotates as a rigid body, the tangential and rolling displacement vector components that are parallel to $\mathbf{n}$ are subtracted at each time step, and scaled to preserve their magnitude.
The velocity of each of the 4 modes $\mathbf{v}_{m}$ is relative to the contact vector and are defined as: 
\begin{subequations}
\begin{align}
\mathbf{v}_{n} =& \left((\mathbf{v}_j - \mathbf{v}_i) \cdot \mathbf{n}\right) \mathbf{n} \label{eqn:vn}\\
\mathbf{v}_{s} =& (\mathbf{v}_j - \mathbf{v}_i) - ((\mathbf{v}_j - \mathbf{v}_i) \cdot \mathbf{n})\mathbf{n} - (R_i\mathbf{\Omega}_i + R_j\mathbf{\Omega}_j) \times \mathbf{n} \label{eqn:vs}\\
\mathbf{v}_r =& -R_{\text{eff}}(\mathbf{\Omega}_i - \mathbf{\Omega}_j) \times \mathbf{n} \label{eqn:vr}\\ 
v_t =& (\mathbf{\Omega}_i - \mathbf{\Omega}_j) \cdot \mathbf{n} \label{eqn:vt}
\end{align}
\end{subequations}
\noindent where $\mathbf{v}_i $ and $\mathbf{\Omega}_i$ are the translational and rotational velocities, respectively. The twisting velocity is a scalar because it is one component of the rotational degrees of freedom.  Fig.~\ref{fig:Modelfig} visualizes the three modes of friction and the associated velocities, forces and torques.

\footnotetext[1]{To use this interaction model in LAMMPS~\cite{StevePlimton1995}, use the following commands: \texttt{pair\_style granular} followed by \texttt{pair\_coeff * * hooke 1 0.5 damping mass\_velocity tangential linear\_history 1 1 $\mu_s$ rolling sds 1 0.5 $\mu_r$ twisting sds 1 0.5 $\mu_t$}. See LAMMPS documentation at lammps.sandia.gov for more details.}

The assumption of linear elastic behavior for inter-particle contacts is reasonably accurate as a model for sufficiently stiff particles.  Note that as an upper limit, for example, glass has a yield stress $\sigma_y \approx 70 $ MPa and would be expected to yield/fracture/fragment, deviating significantly from spherical shape, for $P_a >> 10^{-3} \frac{k_n}{d}$.  Simulations were run at $P_a = 10^{-4} \frac{k_n}{d}$ where $d$ is the diameter.

In all simulations, particles have the same radius $R_i=R=0.5$ and mass $m_i=m=1$. The particle spring and damping parameters are set equal to each other $k_n = k_s = k_r = k_t = 1.0~m/\tau^2$ and $\gamma_n = \gamma_s = \gamma_r = \gamma_t = 0.5~\tau^{-1}$ where $m$ is the particle mass and $\tau = ~\sqrt{m/k_n}$~\cite{Note1}. 
The model parameters and coefficients of rolling and twisting friction used in this study are not based on a specific material. However, the chosen parameters are within relevant values, determined by simulations and a contact mechanics analysis. The contact mechanics analysis and details of the two-overlapping-spheres simulations are in Appendix~\ref{sec:appendix}.
Changes within an order of magnitude of $k_m$ and $\gamma_m$ did not yield qualitative changes in the packing behavior.  If $k_r$ or $k_t$ are orders of magnitude lower than $k_n$ and the normal force is small, the torque resistances are negligible. The key parameters varied in the analysis presented here of packings are the coefficients of the different friction modes, not the spring and damping coefficients.  
A wide range of friction coefficients are studied that include and go above values for typical materials.
For example, copper, bronze and steel spheres have coefficients of rolling resistance $\mu_r$ of $10^{-4}$ to $10^{-2}$~\cite{Tabor1955,Halling1959}, while viscoelastic materials have values of $10^{-3}$ to $10^{-2}$~\cite{Carbone2013}.  In this study, $\mu_{r,t}$ varies from 0 to 100 because of precedent set by previous simulation studies~\cite{Mort2015,Shundyak2007,Silbert2010} and to understand the range of impact of rolling twisting friction for this model.

\subsection{Constant pressure packing simulations}\label{sec:boxeom}
The contact model described in Sec.~\ref{sec:model} was used to perform discrete element, particle-based simulations in LAMMPS~\cite{StevePlimton1995} by integrating Newton's second law with the velocity-Verlet integration scheme.
The particle positions and orientations are updated based on the inter-particle forces $\mathbf{F}_i$ and torques $\boldsymbol{\tau}_i$ calculated by Equations (\ref{eqn:Ftotal}) and (\ref{eqn:Ttotal}). The equations of motion include the degrees of freedom for a deforming box to simulate granular particles under a constant applied pressure tensor. The granular particles are placed within a fully periodic three-dimensional box which is able to change shape with triclinic deformations to maintain the applied pressure tensor. 
A barostat was used in the $N\mathbf{P}_aH$ ensemble to integrate the positions and momenta of the particles and box, where $N$ is the number of particles, $\mathbf{P}_a$ is the applied pressure tensor and $H$ is the enthalpy. The Shinoda-Shiga-Mikami~\cite{Shinoda2004} formulation used in this study combines the hydrostatic equations of Martyna et al.~\cite{Martyna1994} with the strain energy proposed by Parrinello and Rahman~\cite{Parrinello1981}. 

For each pressure and friction state measured, 6 packings of $N=10^4$ diameter $d=1$ non-overlapping particles were generated.  
Simulations were initialized with particles at random positions and low volume fraction $\phi_0 = 0.05$. 
The initial volume fraction $\phi_0$ does not affect the properties of the final packing studied here, so long as $\phi_0$ is not too near, nor above, the jamming volume fraction ($\phi_0 < \phi_{\text{jam}} - 0.3$). 
Initial particle translational and rotational velocities were set to zero. 
The simulation time step was set to $\delta t=0.02\tau$.  A time step of $0.002\tau$ did not change the results for the systems studied within the uncertainties.  

\footnotetext[2]{To apply this symmetric pressure tensors in LAMMPS~\cite{StevePlimton1995}, use \texttt{fix 1 all nph/sphere x 1e-4 1e-4 2.25 y 1e-4 1e-4 2.25 z 1e-4 1e-4 2.25 xy 0.0 0.0 2.25 yz 0.0 0.0 2.25 nreset 1 pchain 0}. See LAMMPS documentation at lammps.sandia.gov for more details.}
After initialization, the particles are isotropically compressed.
The packing method begins with a system at $\phi_0 = 0.05$ and $P=0$, then at $t=0$ a constant pressure $P_{a}=10^{-4}~\frac{k_n}{d}$ with a pressure damping of $P_{\text{damp}}=2.25~\tau^{-1}$ is applied until the system jams. 
Simulations were run for $t/\tau = 10^6$, well above the time to jam $t_{jam}\sim 10^{4} \tau$, defined as the inflection point in the kinetic energy.
The applied symmetric pressure tensor $\mathbf{P}_a$ is defined as: $P_a = P_{a,xx} = P_{a,yy} = P_{a,zz} = 10^{-4}~\frac{k_n}{d}$ and $P_{a,xy} = P_{a,xz} = P_{a,yz} = 0$~\cite{Note2}.
The applied and measured pressure tensors equal each other exactly $\mathbf{P_{\text{int}}}=\mathbf{P_{a}}$ at jamming.

The volume fraction $\phi$ and coordination number $Z$ are the key parameters measured in this study. 
These properties were averaged over the 6 packings generated from the final simulation configurations. 
Measured values of $Z$ are calculated without ``rattlers'', particles that have too few contacts to contribute to the mechanical stability of the packings. Particles were classified as rattlers if $Z_i \le N^c_{\text{ratt}}$, where $Z_i$ is the number of contacts of particle $i$ and $N^c_{\text{ratt}}$ is a friction-dependent minimum number of constraints on the degrees of freedom of motion. 
Each mode of friction contributes constraints to $N^c_{\text{ratt}}$. The friction contributions are determined with constraint counting (see Sec.~\ref{sec:constraint_counting}), and the values of $N^c_{\text{ratt}} = N^c/2$, where $N^c$ is the value in Table~\ref{tab:constraints}. 
An intermediate friction of $\mu_m^c=0.01$ was used to determine whether the sliding, rolling and/or twisting friction mode $m$ would contribute to $N^c_{\text{ratt}}$ which determines the row in Table~\ref{tab:constraints}. The value of $\mu_m^c=0.01$ was chosen because it is the point where friction has an appreciable impact on $\phi$ and $Z$. Other choices of $\mu_m^c$ did not lead to large changes. 
Rattlers are identified iteratively, so that the number of contacts per particle decreases based on the number of rattlers in contact with the particle.  If the number of contacts decreases enough to constitute a rattler, by removing neighboring rattlers, it is counted as such.

\section{Results} \label{sec:results}
\begin{figure}[!htbp]
\centering
\includegraphics[width=1\columnwidth]{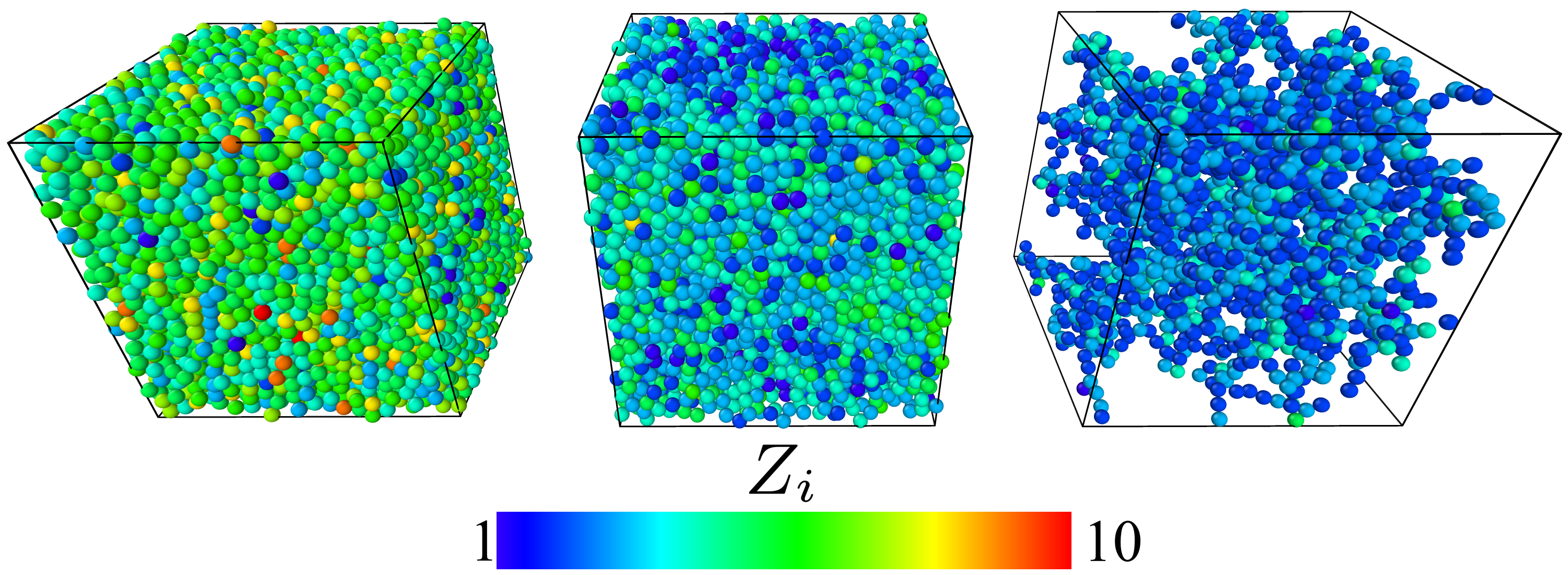}
\caption{Configurations of mechanically stable granular particle packings at three different sliding $s$, rolling $r$ and twisting $t$ friction states: low ($\mu_s=\mu_r=\mu_t = 0$, left), intermediate ($\mu_s=\mu_r=\mu_t = 0.3$, middle) and high ($\mu_s=\mu_r=\mu_t = 1$, right). Rattlers have been removed from the visualization, and each granular particle is colored based on the number of local contacts $Z_i$.}
\label{fig:configs}
\end{figure}

Representative packed granular particle configurations which demonstrate that the inclusion of sliding, rolling and twisting friction causes major microstructural change are shown in Fig.~\ref{fig:configs}.  
Though each system has 10,000 particles, the volume of each is slighty different. Rattlers, which are particles that do not have enough contacts to be mechanically stable because they can move within a mechanically stable packed system, are not visualized (see Sec.~\ref{sec:boxeom} for rattler identification details). The fraction of rattlers increases with the friction coefficient and the number of friction modes.
As the sliding, rolling and twisting friction coefficients increase, particles are more likely to have fewer contacts -- see the color change of particles in Fig.~\ref{fig:configs}. The structure and small fraction of non-rattlers at high sliding $\mu_s$, rolling $\mu_r$ and twisting $\mu_t$ frictions differ considerably from the case where $\mu_s=\mu_r=\mu_t=0$ or even $\mu_s=\mu_r=\mu_t=0.3$. The drastic decrease in the average number of contacts per particle with multiple friction modes can be predicted by balancing contact forces and constraints for a mechanically stable packing.

\subsection{Constraint counting}\label{sec:constraint_counting}
Constraint counting models a packing as a state where the total number of forces and torques on the granular particles equals the total number of constraining contacts to satisfy Maxwell's rigidity criterion~\cite{Maxwell1865}.
For $N$ granular particles in $d$ dimensions there are $N$ normal forces and, if there is friction, $(d-1)N$ tangential forces, $N$ torques for $d=2$, and $3N$ torques for $d=3$.
For mechanical stability, these forces and torques must be balanced by the total \textit{constraining} contacts $NZ$, where $Z$ is the average number of contacts per particle.  The analysis here assumes that any contact, no matter how close it is to the friction limit, is constraining.
Each constraint mode $m$ contributes to the number of constraints per contact, $N^c = \Sigma_m N^c_m$.
The normal contact force law of a hard sphere contributes $N^c_n$ = 1, although for real or simulated hard granular particles, without adhesion, it is only constrained in the repulsive direction. The number of constraints per contact for the other modes in 3D are  $N^c_s=2$, $N^c_r=2$ and $N^c_t=1$.  In 2D $N^c_s=1$, $N^c_r=1$ and $N^c_t=0$.  The total number of constraints and equations $N^{\text{eqn}}$ are then set to equal each other, so that:

\begin{subequations}
\begin{align}
N^{\text{eqn}}=& \begin{cases}
6N,\text{if 3D, frictional}\\
3N,\text{if 2D, frictional}\\
3N,\text{if 3D, frictionless}\\
2N,\text{if 2D, frictionless}
\end{cases} \\
N^{\text{eqn}} =& \frac{N^c}{2}N\left\langle Z\right\rangle\\
\left\langle Z\right\rangle =& \frac{2N^{\text{eqn}}}{N^cN}
\label{eqn:constraints}
\end{align}
\end{subequations}
since a packing has to balance $dN$ forces and $3N$ (3D) or $1N$ (2D) torques for frictional particles. The number of local constraints $N_c^{\text{local}}$, which are used to identify rattlers in the simulation configuration analysis is $N_c^{\text{local}} = N_c/2$.
Table~\ref{tab:constraints} lists the $N_c$ and $Z$ calculated from constraint counting using Equation (\ref{eqn:constraints}). The low value of $Z=2$ when all modes of friction are constraining was previously iocalculated by Liu et al.~\cite{Liu2017b}. The predictions for $Z$ with different modes are compared to simulation results in Sec.~\ref{sec:frictions}. 
\begin{table}[]
\centering
\begin{tabular}{ccc|ll|ll}
	\hline\hline
	\multicolumn{3}{c|}{friction}&\multicolumn{2}{c|}{3D}&\multicolumn{2}{c}{2D}\\
	sliding & rolling & twisting & $N^c$ & $Z$ & $N^c$ & $Z$ \\
	\hline
	n & n & n & 1 & 6 & 1 & 4  \\
	y & n & n & 3 & 4 & 2 & 3  \\
	n & y & n & 3 & 4 & 2 & 3  \\
	n & n & y & 2 & 6 & - & -  \\
	y & y & n & 5 & 12/5 & 3 & 2  \\
	y & n & y & 4 & 3 & - & -  \\
	n & y & y & 4 & 3 & - & -  \\
	y & y & y & 6 & 2 & - & -  \\
	\hline\hline
\end{tabular}
\caption{The average number of contacts per particle $Z$ needed to satisfy the number of constraints per contact $N_c$ for a mechanically stale packing due to sliding, rolling, twisting and the various friction combinations for three-dimensional and two-dimensional particles. The inclusion `y' or exclusion `n' of a friction mode determines $N^c$. Two-dimensional particles do not have the twisting mode and values are omitted accordingly.}
\label{tab:constraints}
\end{table}

\subsection{Packing structure with rolling and twisting}\label{sec:frictions}
\begin{figure}[!htbp]
\centering
\includegraphics[width=\columnwidth]{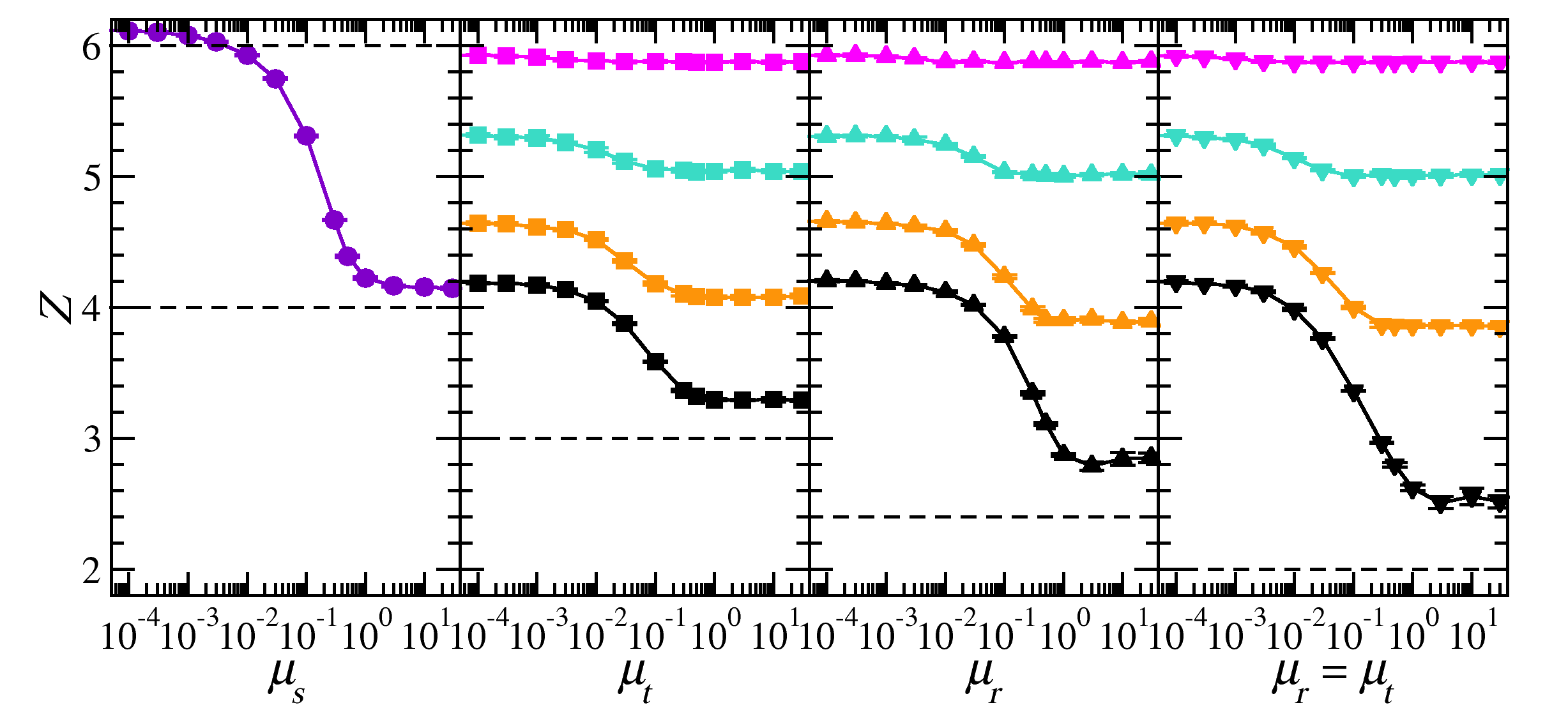}
\caption{Average coordination number without rattlers $Z$ at jamming as a function of sliding ($\mu_s$, far left circles), twisting ($\mu_t$, for different $\mu_s$ where $\mu_r = 0$, center left squares), rolling ($\mu_r$, for different $\mu_s$ where $\mu_t = 0$, center right upward-facing triangles) frictions and where rolling and twisting frictions are set equal to each other (for different $\mu_s$ where $\mu_r$ = $\mu_t$, far right downward-facing triangles).  The leftmost panel shows how $Z$ behaves with $\mu_r=\mu_t$ = 0.0 (violet). For the other panels, data with different sliding frictions are drawn with the following distinct colors going from low to high: $\mu_s$ = 0.01, (magenta),  0.1 (turquoise), 0.3 (orange) and 1 (black).  Constraint counting values (black dashed lines) correspond with the cases shown in Table~\ref{tab:constraints}. Packings are generated at $P_a=10^{-4}~\frac{k_n}{d}$ with $P_{\text{damp}}=2.25~\tau^{-1}$.  Uncertainties are similar in size to the symbols, and solid lines are guides for the eye.}
\label{fig:ZRollTwist}
\end{figure}

Even though constraint counting predicts that rolling and twisting resistances cause large changes in $Z$, simulations often ignore resistance to rolling and twisting.  
Simulations of many particles with the isotropic compression method described in Sec.~\ref{sec:boxeom} can test the constraint counting predictions, and compare with experimental measurements of mechanically stable packing.
Mechanically stable packings were generated with a constant pressure tensor, where diagonal components are set to $P_a=10^{-4}~\frac{k_n}{d}$ and off-diagonal components are set to zero, applied to an initially very dilute system, see Sec.~\ref{sec:boxeom} for more details. The pressure damping for all components is $P_{\text{damp}}=2.25~\tau^{-1}$.
Fig.~\ref{fig:ZRollTwist} quantifies the impact of $\mu_r$ and $\mu_t$ on the coordination number without rattlers seen in Fig.~\ref{fig:configs} and predicted by constraint counting in Table~\ref{tab:constraints}.  
The different panels in Fig.~\ref{fig:ZRollTwist} isolate the impact of each friction mode. For sliding friction without rolling and twisting, $Z$ decreases with increasing $\mu_s$ as observed in previous volume-controlled packings~\cite{Silbert2010}.  As $\mu_s\to 0$, $Z$ approaches the 3D frictionless limit $Z=6$.  As $\mu_s$ increases, $Z$ continuously decreases to $Z=4$, the limit predicted by constraint counting. Constraint counting predicts that Z decreases from 6 to 4 for any nonzero sliding friction. Shundyak et al.~\cite{Shundyak2007} found that, for 2-dimensional particles with sliding friction in the hard-sphere limit, the $Z$ predicted from constraint counting equals the number of contacts minus the mobilized or plastic contacts per particle. Rolling and twisting friction modes have similar effects on $Z$ with some distinctions.

The center two panels of Fig.~\ref{fig:ZRollTwist} show that for low $\mu_s$, $Z$ is insensitive to rolling and twisting friction. The insensitivity to $\mu_r$ and $\mu_t$ at low $\mu_s$ is due to how friction is modeled. The contact point can move and disengage the rolling and twisting torques if the sliding friction is too low.
As sliding friction is increased, $\mu_s > 10^{-2}$, rolling and twisting friction begin to affect $Z$ in a similar way as sliding friction. Increasing sliding friction decreases $Z$ at the low-$\mu_{r,t}$ values as well as increases the magnitude of the impact $\mu_r$ and $\mu_t$ have on $Z$ at high-$\mu_{r,t}$ values. The scale of the decrease in $Z$ depends on how many constraints a friction mode contributes. Because rolling friction contributes two constraints to rotational motion compared to one from twisting friction, rolling friction leads to a larger decrease in magnitude for $Z$. Constraint counting predicts those magnitudes, see Table~\ref{tab:constraints}.
Whereas with only sliding friction $\lim_{\mu_s\to \infty} Z = 4$, the inclusion of twisting friction leads to $\lim_{\mu_{s,t}\to \infty}Z = 3.291\pm0.009$, rolling friction leads to $\lim_{\mu_{s,r}\to \infty}Z = 2.85\pm0.05$ and for all three frictions $\lim_{\mu_{s,r,t}\to \infty}Z = 2.50\pm0.05$ (limiting $Z$ values were taken as the minimum measured, and reported in Fig. \ref{fig:ZRollTwist}). Similar limiting behavior in $Z$ was recently observed in shear jammed dense suspension simulations~\cite{Singh2020}.  Any process that includes granular packings is likely impacted by the large decreases in the average number of contacts per particle from frictionless ($Z=6$) to large sliding, rolling and twisting friction ($Z=2.5$).

These values are close to, but consistently greater than, the values predicted by constraint counting indicated by the dashed horizontal lines. A portion of the under-estimation is because the $Z$ reported in Fig.~\ref{fig:ZRollTwist} is without rattlers. Taking rattlers out decreases the number of particles used to calculate $Z$, without much change in the number of contacts, and is not accounted for in constraint counting.  
To understand the larger constraint counting-simulation $Z$ discrepancy, consider the $Z=2$ prediction for all three friction modes. Because there is no cohesion in this model, a particle with one contact is a rattler.  Therefore, the only way for there to be a mechanically stable system with $Z=2$, is if all non-rattlers have exactly 2 contacts.  A stable packing of particles with only two contacts would be highly unlikely. The present system instead forms packings with a few $Z>2$ particles between chains of $Z=2$  for an average $Z \ge 2.5$. Previous simulations that included cohesion formed packings with $Z=2$~\cite{Liu2017b}, support this explanation.

From the constraint counting predictions of the number of constraints per contact $N^c_m$ and the critical value of mode $m$ friction $\mu_{m,c}$ at which $Z$ is half way between the two limiting cases, $Z_{m,c} = \left(Z(\mu_m \to 0)+Z(\mu_m \to \infty)\right)/2$, the $Z$ behavior can be predicted without simulation data as:
\begin{eqnarray}
&Z&= N^{\text{eqn}} - \tanh\left(\frac{\mu_s}{\mu_{s,c}}\right) \nonumber\\ 
&\times&\left[N^{\text{c}}_s\frac{}{}+ N^{\text{c}}_r \tanh\left(\frac{\mu_r}{\mu_{r,c}}\right)+N^{\text{c}}_t \tanh\left(\frac{\mu_t}{\mu_{t,c}}\right)\right]
\label{eqn:Zfit}
\end{eqnarray}
where the number of equations $N^{\text{eqn}}$ and constraints $N^{c}_m$ are detailed in Table~\ref{tab:constraints}.  For this model, $\mu_{s,c} \simeq \mu_{r,c} \simeq \mu_{t,c} \simeq 0.3$, and thus $Z = 6 - \tanh\left(\frac{\mu_s}{0.3}\right) \left[2+2\tanh\left(\frac{\mu_r}{0.3}\right)+ \tanh\left(\frac{\mu_t}{0.3}\right)\right]$. The $\tanh$ function is chosen because it matches the correct limiting behavior and exponentially connects the limits.  The sliding friction term, $\tanh\left(\frac{\mu_s}{\mu_{s,c}}\right)$ multiplies the rolling and twisting terms because no resistance to the sliding mode can lead granular particles to loose contact.  The rolling and twisting modes of rotational motion cannot individually lead to contact disengagement. Because Equation (\ref{eqn:Zfit}) is informed by constraint counting, its limits of $Z$ are those predicted by constraint counting and do not match the simulation results. Equation (\ref{eqn:Zfit}) is a tool to estimate $Z$ if the sliding, rolling and twisting friction coefficients are known, and could be used to select a material or model with desired packing properties.

\begin{figure}[!htbp]
\centering
\includegraphics[width=\columnwidth]{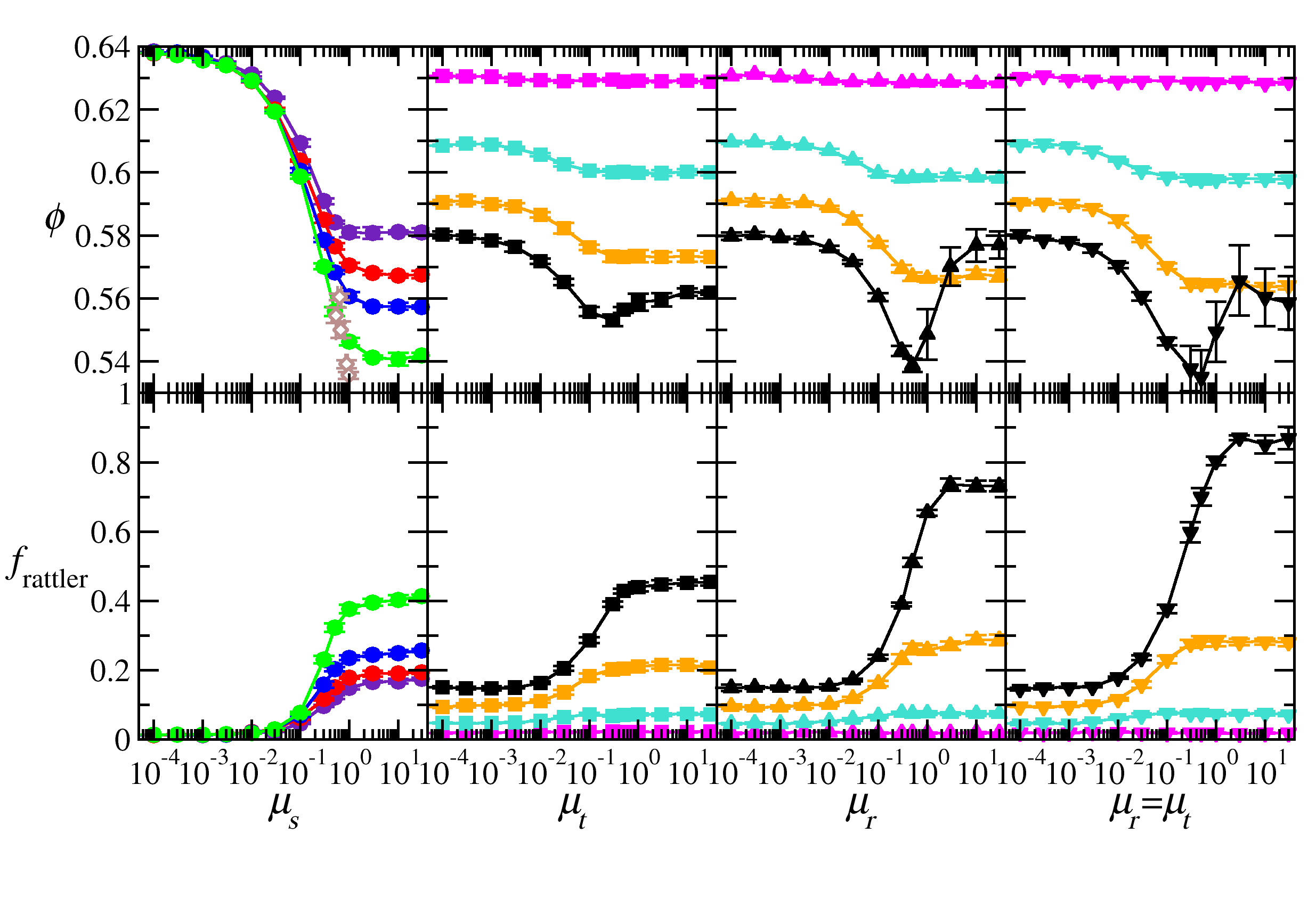}
\caption{Volume fraction ($\phi$, top) and fraction of rattlers ($f_{\text{rattler}}$, bottom) at jamming as a function of sliding ($\mu_s$, for different $\mu_t = \mu_r$, far left circles),  twisting ($\mu_t$, for different $\mu_s$ where $\mu_r = 0$, center left squares), rolling ($\mu_r$, for different $\mu_s$ where $\mu_t = 0$, center right upward-facing triangles) frictions and where rolling and twisting friction coefficients are equal (for different $\mu_s$ where $\mu_r$ = $\mu_t$, far right downward-facing triangles).  The leftmost panel shows a series of curves that represent different rolling and twisting friction coefficients: $\mu_r=\mu_t$ = 0.0, (violet), $\mu_r=\mu_t$ = 0.01, (red), $\mu_r=\mu_t$ = 0.03, (blue) and $\mu_r=\mu_t$ = 0.1 (green).  For $\mu_r,\mu_t>$0.1, there is little change in $\phi(\mu_s)$ behavior.  Experimental $\phi$ values of Farerll et al.~\cite{Farrell2010} for different materials with associated $\mu_s$ are shown as brown diamonds. For the other panels, the colors are the same as Fig.~\ref{fig:ZRollTwist}.
}
\label{fig:PhiRollTwist} 
\end{figure}

The distribution of coordination numbers in high and low $Z$ packings are visualized in Fig.~\ref{fig:configs}.
To be mechanically stable with so few contacts, a large fraction of the granular particles must be rattlers.
Not only are there fewer non-rattler particles, but the distribution of non-rattlers is very heterogeneous. 
Fig.~\ref{fig:PhiRollTwist} quantifies the fraction of rattlers as a function of the various friction modes. The rattler fraction increases monotonically with the friction coefficient of each mode.  $Z$ and $f_{\text{rattler}}$ transitions from the low-friction to high-friction limits are similar. Rattlers become the majority with large friction $\mu_{r,t}$ values if $\mu_s > 0.3$. Such microstructure must be very fragile (quantification of packing strength is subject of future study).

Fig.~\ref{fig:PhiRollTwist} includes the volume fraction dependence on the different friction modes. $Z$ and $\phi$ behave similarly, except for a minimum in $\phi$ for high $\mu_s$. The minimum in $\phi$ is likely due to contacts saturating at the Coulomb friction criteria.  Once constraining contacts saturate, the contacts can slide to form a denser packing, while maintaining their network. As seen with $Z$ and $f_{\text{rattler}}$, sliding, rolling and twisting friction cause a larger decrease in packing fraction.  

Fig.~\ref{fig:PhiRollTwist} also shows the volume fraction of the packings generated by experiments. Experimental values are from Farrell et al.~\cite{Farrell2010} and were performed by slowly settling spheres of different sliding friction coefficients, set by the material (steel, aluminum, acrylic or teflon). The experimental $\phi$ values are considerably below the simulation $\phi$ values without rolling and twisting friction. As seen in Fig.~\ref{fig:PhiRollTwist}, there is agreement with experiment for $\phi$ only with rolling and twisting friction, using the compression method described in Sec.~\ref{sec:boxeom} with $P_a = 10^{-4}~\frac{k_n}{d}$.  To match experimental values for various acrylics, teflon and steel, $\mu_r=\mu_t=0.1$ is required, while for aluminum our results for $\mu_r=\mu_t=0.03$ match the experiment. Those values required to match experiments are similar to the value of $\mu_r=0.07$ used in recent dense suspension simulations to match experimental shear viscosities~\cite{Singh2020}.  
Not only do rolling and twisting friction have a major impact on microstructure, as measured by $Z$, but they should be included in experimentally-relevant particle models. 

\section{Conclusion}\label{sec:conclusion}
Discrete-element, particle based simulations of 3-dimensional granular particles demonstrated that rolling and twisting friction leads to large microstructural changes in mechanically stable packings, as insinuated by constraint counting. Agreement with experimental volume fractions was only attained with rolling and twisting friction on ($\mu_r=\mu_t=0.1$) using this simulation protocol. These loose packings, $\phi=0.53$ demonstrated the importance of different friction modes in real granular systems. The pressure-controlled compression protocol generated very loose packings with less computational effort than other methods. 
A deeper investigation into pressure-controlled simulation packing methods (compression, relaxation and tapping) and parameters (pressure, pressure damping and box drag) is the subject of a forthcoming article. 

The decrease in the coordination number $Z$ was predicted from constraint counting -- both rolling and twisting friction impose extra constraints per contact. The decrease of the volume fraction and coordination number from low- to high-friction values were gradual for all three friction modes, as observed for sliding friction.  Unlike sliding friction, the impact of resistance to rolling and twisting depended on the sliding resistance magnitude. When multiple friction modes were included, such as sliding with rolling or twisting, the coordination number predicted by constraint counting is considerably less than the value measured from simulations.  Nonetheless, for very high friction, $\mu_s=\mu_r=\mu_t=1$, a jammed system with $Z=2.5$ was observed. The majority of particles were rattlers in systems with such low coordination numbers, generally if $Z<3$. Based on the knowledge gained from this information, an expression to predict $Z(\mu_s,\mu_r,\mu_t)$ was proposed to aid constitutive models and future parametric studies.                                                               

The effect of rolling and twisting friction on packing illustrates the importance of including those modes to match experimental results and offers insight into the magnitude of those frictions required to induce property changes for other granular systems.
Future work will focus on the impact of rolling and twisting friction on rheology and the material strength of packings. The publicly available rolling and twisting interaction models in LAMMPS enable the presented and future work. 

\begin{acknowledgments}
This work was performed, in part, at the Center for Integrated Nanotechnologies, an Office of Science User Facility operated for the U.S. Department of Energy (DOE) Office of Science. 
Sandia National Laboratories is a multi-mission laboratory managed and operated by National Technology and Engineering Solutions of Sandia, LLC., a wholly owned subsidiary of Honeywell International, Inc., for the U.S. DOE’s National Nuclear Security Administration under contract DE-NA-0003525. 
The views expressed in the article do not necessarily represent the views of the U.S. DOE or the United States Government.
\end{acknowledgments}

\appendix
\section{Rolling and twisting parameterization}\label{sec:appendix}
Rolling and twisting resistance depend on the normal and tangential forces in most contact mechanics models~\cite{Johnson1985}. Although simple, the spring-dashpot-slider system enables the study of each mode independently.  This model allows for resistance to ``pure'' rolling. However, if there is no sliding friction contacts can relax the rolling resistance by sliding off the contact~\cite{Khan2005,Jiang2006,Zhu2006,Ai2011,Wang2015c}.

Sources of twisting and sliding resistance are essentially the same -- twisting friction is largely generated by rotational, versus translational, displacements within the contact area~\cite{Johnson1985}.  
Therefore, twisting resistance should be related both to the tangential mode, from the sliding friction coefficient, and to the normal mode, from the contact area. 
There are more identified sources for rolling than twisting resistance. Micro-slip at the interface, inelastic deformation and surface roughness all lead to resistance because of the pressure difference between the leading and trailing ends of the rolling contact~\cite{Johnson1985}. 
Micro-slip, related to incipient sliding~\cite{Johnson1985}, occurs from creep of the interfacing material and the difference in shear forces at the interface when material in the contact area slips. Those phenomena arise from differences in material elastic constants, curvature and torsion. The difference of strain on either side of the rolling contact causes inelastic deformation, another source of rolling resistance~\cite{Eldredge1955,Tabor1955}.  
In most cases, inelastic deformation creates the largest rolling resistance contribution, and can be characterized experimentally by a ``hysteresis loss factor''~\cite{Johnson1985}. 
Surface roughness and viscoelasticity can lead to rolling resistance, although likely at a lower magnitude than inelastic deformation and micro-slip.
Surface roughness induces changes in the center-of-mass separation of the two bodies~\cite{Drutowski1959}, and increases the real contact pressure~\cite{Halling1959}.  Viscoelastic materials have a velocity-dependent rolling resistance, because of the balance in relaxation and observation times in rolling resistance for viscoelastic materials~\cite{May1959,Flom1959,Hunter1960}.  Approximations for the relative magnitude of rolling resistance have been made in the particle-based simulation literature.

Previous particle simulation studies that used spring interaction models made approximations for how the rolling spring constant $k_r$ relates to more common parameters.
Iwashita and Oda~\cite{Iwashita1998} set $k_r = k_s/R^2$ by equating first-order approximations of the shear and rolling elastic displacement.
Jiang et al.~\cite{Jiang2005} instead assumed $k_r = \frac{1}{12}k_n r_a^2$ and $\gamma_r =  \frac{1}{12}\gamma_n r_a^2$, from Hertzian contact theory by representing a rolling contact as springs in parallel, where $r_a$ is the contact radius and is calculated for each contact.  
A more exact analytical solution for a viscous sphere on a hard plane results in a relatively small value for $k_r$ that is friction dependent and therefore does not work for our model~\cite{Brilliantov1998}. 
Here, we perform our own analysis of contact mechanics models to approximate the parameters of the twisting and rolling pseudo-forces.

The twisting resistive moment in Hertzian theory is: $F_t = \frac{16}{3}G r_a^3\theta_t$ where $\theta_t$ is the twisting angle, $G$ is the shear modulus and $r_a$ is the contact radius~\cite{Lubkin1951}. This twisting model is associated with ``no slip'' which yields a linear model with $\theta_t$.  
The contact radius can be estimated from the normal force as $r_a = \left( \frac{3F_nR(1-\nu^2)}{2E} \right)^{1/3}$, where $R$ is the particle radius, $\nu$ is the Poisson ratio and $E$ is the elastic modulus~\cite{Johnson1985}.
By using the Hertzian approximation for $r_a$ in the Lubkin twisting force $F_t$  theory and by inserting $k_t\theta_t$ and $k_n\delta_t$ for tangential and normal Hookean-spring contact models, the following relationship for the twisting spring constant is found: $k_t/k_n \propto \frac{8GR(1-\nu^2)}{E}$. 
For steel $G = 79.3$ GPa, $E = 200$ GPa, $\nu = 0.27$ and therefore $k_t/k_n \simeq 1.5$.  For rubber $G = 0.0003$ GPa, $E = 0.001$ GPa, $\nu = 0.5$ and therefore $k_t/k_n \simeq 0.9$.  
A similar analysis can also be carried out for the rolling pseudo-force.
Based on Johnson's formulation of elastic deformation~\cite{Johnson1958a,Johnson1958}, the moment due to elastic creep between spheres is $\mathbf{F}_r= \frac{32(2-\nu)}{9(3-2\nu)}Gr_a^3\mathbf{v}_r$.  
Using the same assumption for the contact radius of two elastic bodies used for twisting, we get $\mathbf{F}_r= \frac{16(2-\nu)(1-\nu^2)}{3(3-2\nu)}\frac{G}{E}R\mathbf{F}_n\theta_r$. 
Since we model $F_r=k_r\theta_r$ and $F_n=k_n\delta_n$, the ratio of rolling and normal forces scales as $k_r/k_n \propto  \frac{16(2-\nu)(1-\nu^2)}{3(3-2\nu)}\frac{G}{E}R$, for elastic deformation.  For steel $k_r/k_n \simeq 0.7$, and for rubber $k_r/k_n \simeq 0.5$.

To empirically identify realistic parameters for the rolling and twisting modes, DEM simulations of simple configurations were performed. Two overlapping suspended spheres, with fixed translational positions (and fixed $\mathbf{F}_n$), were given an initial relative rolling or twisting angular velocity.  Nonphysical values of $k_m$ and $\gamma_m$ gave long-lived oscillations and/or over-damped decay of the torque.
The rolling spring-dashpot-slider has realistic behavior for $0 < k_r/\gamma_r < 1$, yet $ k_r/\gamma_r > 1$ is realistic if $k_n/k_r < 10$.  Realistic twisting angle and torque behavior was found for $0 < k_t/\gamma_t \le 2$.  

\bibliography{GranularPackingFrictionModes.bib}

\begin{thebibliography}{61}%
\makeatletter
\providecommand \@ifxundefined [1]{%
 \@ifx{#1\undefined}
}%
\providecommand \@ifnum [1]{%
 \ifnum #1\expandafter \@firstoftwo
 \else \expandafter \@secondoftwo
 \fi
}%
\providecommand \@ifx [1]{%
 \ifx #1\expandafter \@firstoftwo
 \else \expandafter \@secondoftwo
 \fi
}%
\providecommand \natexlab [1]{#1}%
\providecommand \enquote  [1]{``#1''}%
\providecommand \bibnamefont  [1]{#1}%
\providecommand \bibfnamefont [1]{#1}%
\providecommand \citenamefont [1]{#1}%
\providecommand \href@noop [0]{\@secondoftwo}%
\providecommand \href [0]{\begingroup \@sanitize@url \@href}%
\providecommand \@href[1]{\@@startlink{#1}\@@href}%
\providecommand \@@href[1]{\endgroup#1\@@endlink}%
\providecommand \@sanitize@url [0]{\catcode `\\12\catcode `\$12\catcode
  `\&12\catcode `\#12\catcode `\^12\catcode `\_12\catcode `\%12\relax}%
\providecommand \@@startlink[1]{}%
\providecommand \@@endlink[0]{}%
\providecommand \url  [0]{\begingroup\@sanitize@url \@url }%
\providecommand \@url [1]{\endgroup\@href {#1}{\urlprefix }}%
\providecommand \urlprefix  [0]{URL }%
\providecommand \Eprint [0]{\href }%
\providecommand \doibase [0]{https://doi.org/}%
\providecommand \selectlanguage [0]{\@gobble}%
\providecommand \bibinfo  [0]{\@secondoftwo}%
\providecommand \bibfield  [0]{\@secondoftwo}%
\providecommand \translation [1]{[#1]}%
\providecommand \BibitemOpen [0]{}%
\providecommand \bibitemStop [0]{}%
\providecommand \bibitemNoStop [0]{.\EOS\space}%
\providecommand \EOS [0]{\spacefactor3000\relax}%
\providecommand \BibitemShut  [1]{\csname bibitem#1\endcsname}%
\let\auto@bib@innerbib\@empty
\bibitem [{\citenamefont {Skinner}(1969)}]{Skinner1969}%
  \BibitemOpen
  \bibfield  {author} {\bibinfo {author} {\bibfnamefont {A.~E.}\ \bibnamefont
  {Skinner}},\ }\bibfield  {title} {\bibinfo {title} {{A note on influence of
  interparticle friction on shearing strength of a random assembly of spherical
  particles}},\ }\href@noop {} {\bibfield  {journal} {\bibinfo  {journal}
  {Geotechnique}\ }\textbf {\bibinfo {volume} {19}},\ \bibinfo {pages} {150}
  (\bibinfo {year} {1969})}\BibitemShut {NoStop}%
\bibitem [{\citenamefont {Mort}\ \emph {et~al.}(2015)\citenamefont {Mort},
  \citenamefont {Michaels}, \citenamefont {Behringer}, \citenamefont
  {Campbell}, \citenamefont {Kondic}, \citenamefont {{Kheiripour Langroudi}},
  \citenamefont {Shattuck}, \citenamefont {Tang}, \citenamefont {Tardos},\ and\
  \citenamefont {Wassgren}}]{Mort2015}%
  \BibitemOpen
  \bibfield  {author} {\bibinfo {author} {\bibfnamefont {P.}~\bibnamefont
  {Mort}}, \bibinfo {author} {\bibfnamefont {J.~N.}\ \bibnamefont {Michaels}},
  \bibinfo {author} {\bibfnamefont {R.~P.}\ \bibnamefont {Behringer}}, \bibinfo
  {author} {\bibfnamefont {C.~S.}\ \bibnamefont {Campbell}}, \bibinfo {author}
  {\bibfnamefont {L.}~\bibnamefont {Kondic}}, \bibinfo {author} {\bibfnamefont
  {M.}~\bibnamefont {{Kheiripour Langroudi}}}, \bibinfo {author} {\bibfnamefont
  {M.}~\bibnamefont {Shattuck}}, \bibinfo {author} {\bibfnamefont
  {J.}~\bibnamefont {Tang}}, \bibinfo {author} {\bibfnamefont {G.~I.}\
  \bibnamefont {Tardos}},\ and\ \bibinfo {author} {\bibfnamefont
  {C.}~\bibnamefont {Wassgren}},\ }\bibfield  {title} {\bibinfo {title} {{Dense
  granular flow - A collaborative study}},\ }\href
  {https://doi.org/10.1016/j.powtec.2015.06.006} {\bibfield  {journal}
  {\bibinfo  {journal} {Powder Technol.}\ }\textbf {\bibinfo {volume} {284}},\
  \bibinfo {pages} {571} (\bibinfo {year} {2015})}\BibitemShut {NoStop}%
\bibitem [{\citenamefont {Singh}\ \emph {et~al.}(2020)\citenamefont {Singh},
  \citenamefont {Ness}, \citenamefont {Seto}, \citenamefont {de~Pablo},\ and\
  \citenamefont {Jaeger}}]{Singh2020}%
  \BibitemOpen
  \bibfield  {author} {\bibinfo {author} {\bibfnamefont {A.}~\bibnamefont
  {Singh}}, \bibinfo {author} {\bibfnamefont {C.}~\bibnamefont {Ness}},
  \bibinfo {author} {\bibfnamefont {R.}~\bibnamefont {Seto}}, \bibinfo {author}
  {\bibfnamefont {J.~J.}\ \bibnamefont {de~Pablo}},\ and\ \bibinfo {author}
  {\bibfnamefont {H.~M.}\ \bibnamefont {Jaeger}},\ }\bibfield  {title}
  {\bibinfo {title} {{Shear thickening and jamming of dense suspensions: the
  roll of friction}},\ }\href {https://doi.org/10.1103/PhysRevLett.124.248005}
  {\bibfield  {journal} {\bibinfo  {journal} {Phys. Rev. Lett.}\ }\textbf
  {\bibinfo {volume} {124}},\ \bibinfo {pages} {248005} (\bibinfo {year}
  {2020})}\BibitemShut {NoStop}%
\bibitem [{\citenamefont {Estrada}\ \emph {et~al.}(2008)\citenamefont
  {Estrada}, \citenamefont {Taboada},\ and\ \citenamefont
  {Radja{\"{i}}}}]{Estrada2008}%
  \BibitemOpen
  \bibfield  {author} {\bibinfo {author} {\bibfnamefont {N.}~\bibnamefont
  {Estrada}}, \bibinfo {author} {\bibfnamefont {A.}~\bibnamefont {Taboada}},\
  and\ \bibinfo {author} {\bibfnamefont {F.}~\bibnamefont {Radja{\"{i}}}},\
  }\bibfield  {title} {\bibinfo {title} {{Shear strength and force transmission
  in granular media with rolling resistance}},\ }\href
  {https://doi.org/10.1103/PhysRevE.78.021301} {\bibfield  {journal} {\bibinfo
  {journal} {Phys. Rev. E}\ }\textbf {\bibinfo {volume} {78}},\ \bibinfo
  {pages} {021301} (\bibinfo {year} {2008})}\BibitemShut {NoStop}%
\bibitem [{\citenamefont {Liu}\ \emph {et~al.}(2018)\citenamefont {Liu},
  \citenamefont {Liu},\ and\ \citenamefont {Mao}}]{Liu2018e}%
  \BibitemOpen
  \bibfield  {author} {\bibinfo {author} {\bibfnamefont {Y.}~\bibnamefont
  {Liu}}, \bibinfo {author} {\bibfnamefont {H.}~\bibnamefont {Liu}},\ and\
  \bibinfo {author} {\bibfnamefont {H.}~\bibnamefont {Mao}},\ }\bibfield
  {title} {\bibinfo {title} {{The influence of rolling resistance on the
  stress-dilatancy and fabric anisotropy of granular materials}},\ }\href
  {https://doi.org/10.1007/s10035-017-0780-z} {\bibfield  {journal} {\bibinfo
  {journal} {Granular Matter}\ }\textbf {\bibinfo {volume} {20}},\ \bibinfo
  {pages} {12} (\bibinfo {year} {2018})}\BibitemShut {NoStop}%
\bibitem [{\citenamefont {Wu}\ \emph {et~al.}(2019)\citenamefont {Wu},
  \citenamefont {Ma}, \citenamefont {Zhou}, \citenamefont {Wang},\ and\
  \citenamefont {Chang}}]{Wu2019}%
  \BibitemOpen
  \bibfield  {author} {\bibinfo {author} {\bibfnamefont {W.}~\bibnamefont
  {Wu}}, \bibinfo {author} {\bibfnamefont {G.}~\bibnamefont {Ma}}, \bibinfo
  {author} {\bibfnamefont {W.}~\bibnamefont {Zhou}}, \bibinfo {author}
  {\bibfnamefont {D.}~\bibnamefont {Wang}},\ and\ \bibinfo {author}
  {\bibfnamefont {X.}~\bibnamefont {Chang}},\ }\bibfield  {title} {\bibinfo
  {title} {{Force transmission and anisotropic characteristics of sheared
  granular materials with rolling resistance}},\ }\href
  {https://doi.org/10.1007/s10035-019-0938-y} {\bibfield  {journal} {\bibinfo
  {journal} {Granular Matter}\ }\textbf {\bibinfo {volume} {21}},\ \bibinfo
  {pages} {88} (\bibinfo {year} {2019})}\BibitemShut {NoStop}%
\bibitem [{\citenamefont {Mari}\ and\ \citenamefont {Seto}(2019)}]{Mari2019}%
  \BibitemOpen
  \bibfield  {author} {\bibinfo {author} {\bibfnamefont {R.}~\bibnamefont
  {Mari}}\ and\ \bibinfo {author} {\bibfnamefont {R.}~\bibnamefont {Seto}},\
  }\bibfield  {title} {\bibinfo {title} {{Force transmission and the order
  parameter of shear thickening}},\ }\href {https://doi.org/10.1039/c9sm01223k}
  {\bibfield  {journal} {\bibinfo  {journal} {Soft Matter}\ }\textbf {\bibinfo
  {volume} {15}},\ \bibinfo {pages} {6650} (\bibinfo {year}
  {2019})}\BibitemShut {NoStop}%
\bibitem [{\citenamefont {Guy}\ \emph {et~al.}(2018)\citenamefont {Guy},
  \citenamefont {Richards}, \citenamefont {Hodgson}, \citenamefont {Blanco},\
  and\ \citenamefont {Poon}}]{Guy2018}%
  \BibitemOpen
  \bibfield  {author} {\bibinfo {author} {\bibfnamefont {B.~M.}\ \bibnamefont
  {Guy}}, \bibinfo {author} {\bibfnamefont {J.~A.}\ \bibnamefont {Richards}},
  \bibinfo {author} {\bibfnamefont {D.~J.~M.}\ \bibnamefont {Hodgson}},
  \bibinfo {author} {\bibfnamefont {E.}~\bibnamefont {Blanco}},\ and\ \bibinfo
  {author} {\bibfnamefont {W.~C.~K.}\ \bibnamefont {Poon}},\ }\bibfield
  {title} {\bibinfo {title} {{Constraint-Based Approach to Granular Dispersion
  Rheology}},\ }\href {https://doi.org/10.1103/PhysRevLett.121.128001}
  {\bibfield  {journal} {\bibinfo  {journal} {Phys. Rev. Lett.}\ }\textbf
  {\bibinfo {volume} {121}},\ \bibinfo {pages} {128001} (\bibinfo {year}
  {2018})}\BibitemShut {NoStop}%
\bibitem [{\citenamefont {Bardet}(1994)}]{Bardet1994}%
  \BibitemOpen
  \bibfield  {author} {\bibinfo {author} {\bibfnamefont {J.~P.}\ \bibnamefont
  {Bardet}},\ }\bibfield  {title} {\bibinfo {title} {{Observations on the
  effects of particle rotations on the failure of idealized granular
  materials}},\ }\href {https://doi.org/10.1016/0167-6636(94)00006-9}
  {\bibfield  {journal} {\bibinfo  {journal} {Mech. Mater.}\ }\textbf {\bibinfo
  {volume} {18}},\ \bibinfo {pages} {159} (\bibinfo {year} {1994})}\BibitemShut
  {NoStop}%
\bibitem [{\citenamefont {Iwashita}\ and\ \citenamefont
  {Oda}(1998)}]{Iwashita1998}%
  \BibitemOpen
  \bibfield  {author} {\bibinfo {author} {\bibfnamefont {K.}~\bibnamefont
  {Iwashita}}\ and\ \bibinfo {author} {\bibfnamefont {M.}~\bibnamefont {Oda}},\
  }\bibfield  {title} {\bibinfo {title} {{Rolling resistance at contacts in
  simulation of shear band development by DEM}},\ }\href@noop {} {\bibfield
  {journal} {\bibinfo  {journal} {J. Eng. Mech.}\ }\textbf {\bibinfo {volume}
  {124}},\ \bibinfo {pages} {285} (\bibinfo {year} {1998})}\BibitemShut
  {NoStop}%
\bibitem [{\citenamefont {Tordesillas}\ \emph {et~al.}(2005)\citenamefont
  {Tordesillas}, \citenamefont {Peters},\ and\ \citenamefont
  {Muthuswamy}}]{Tordesillas2016}%
  \BibitemOpen
  \bibfield  {author} {\bibinfo {author} {\bibfnamefont {A.}~\bibnamefont
  {Tordesillas}}, \bibinfo {author} {\bibfnamefont {J.}~\bibnamefont
  {Peters}},\ and\ \bibinfo {author} {\bibfnamefont {M.}~\bibnamefont
  {Muthuswamy}},\ }\bibfield  {title} {\bibinfo {title} {{Role of particle
  rotations and rolling resistance in a semi-infinite particulate solid
  indented by a rigid flat punch}},\ }\href
  {https://doi.org/10.21914/anziamj.v46i0.958} {\bibfield  {journal} {\bibinfo
  {journal} {ANZIAM J.}\ }\textbf {\bibinfo {volume} {46}},\ \bibinfo {pages}
  {C260} (\bibinfo {year} {2005})}\BibitemShut {NoStop}%
\bibitem [{\citenamefont {Wang}\ and\ \citenamefont {Mora}(2008)}]{Wang2008a}%
  \BibitemOpen
  \bibfield  {author} {\bibinfo {author} {\bibfnamefont {Y.}~\bibnamefont
  {Wang}}\ and\ \bibinfo {author} {\bibfnamefont {P.}~\bibnamefont {Mora}},\
  }\bibfield  {title} {\bibinfo {title} {{Modeling Wing Crack Extension:
  Implications for the Ingredients of Discrete Element Model}},\ }in\ \href
  {https://doi.org/https://doi.org/10.1007/978-3-7643-8757-0_9} {\emph
  {\bibinfo {booktitle} {Earthquakes: Simulations, Sources and Tsunamis}}},\
  \bibinfo {editor} {edited by\ \bibinfo {editor} {\bibfnamefont
  {K.}~\bibnamefont {Tiampo}}, \bibinfo {editor} {\bibfnamefont
  {D.}~\bibnamefont {Weatherley}},\ and\ \bibinfo {editor} {\bibfnamefont
  {S.}~\bibnamefont {Weinstein}}}\ (\bibinfo  {publisher} {Birkh{\"{a}}user
  Basel},\ \bibinfo {year} {2008})\ pp.\ \bibinfo {pages}
  {609--620}\BibitemShut {NoStop}%
\bibitem [{\citenamefont {Reynolds}(1875)}]{Reynolds1875}%
  \BibitemOpen
  \bibfield  {author} {\bibinfo {author} {\bibfnamefont {O.}~\bibnamefont
  {Reynolds}},\ }\bibfield  {title} {\bibinfo {title} {{On rolling-friction}},\
  }\href@noop {} {\bibfield  {journal} {\bibinfo  {journal} {Phil. Trans. Royal
  Society}\ }\textbf {\bibinfo {volume} {166}},\ \bibinfo {pages} {155}
  (\bibinfo {year} {1875})}\BibitemShut {NoStop}%
\bibitem [{\citenamefont {Hertz}(1882)}]{Hertz1882}%
  \BibitemOpen
  \bibfield  {author} {\bibinfo {author} {\bibfnamefont {H.}~\bibnamefont
  {Hertz}},\ }\bibfield  {title} {\bibinfo {title} {{On the contact of elastic
  solids}},\ }\href@noop {} {\bibfield  {journal} {\bibinfo  {journal} {J.
  reine und angewandte Mathematik}\ }\textbf {\bibinfo {volume} {92}},\
  \bibinfo {pages} {156} (\bibinfo {year} {1882})}\BibitemShut {NoStop}%
\bibitem [{\citenamefont {Brilliantov}\ and\ \citenamefont
  {P{\"{o}}schel}(1998)}]{Brilliantov1998}%
  \BibitemOpen
  \bibfield  {author} {\bibinfo {author} {\bibfnamefont {N.~V.}\ \bibnamefont
  {Brilliantov}}\ and\ \bibinfo {author} {\bibfnamefont {T.}~\bibnamefont
  {P{\"{o}}schel}},\ }\bibfield  {title} {\bibinfo {title} {{Rolling friction
  of a viscous sphere on a hard plane}},\ }\href
  {https://doi.org/10.1209/epl/i1998-00281-7} {\bibfield  {journal} {\bibinfo
  {journal} {Europhys. Lett.}\ }\textbf {\bibinfo {volume} {42}},\ \bibinfo
  {pages} {511} (\bibinfo {year} {1998})}\BibitemShut {NoStop}%
\bibitem [{\citenamefont {P{\"{o}}schel}\ \emph {et~al.}(1999)\citenamefont
  {P{\"{o}}schel}, \citenamefont {Schwager},\ and\ \citenamefont
  {Brilliantov}}]{Poschel1999}%
  \BibitemOpen
  \bibfield  {author} {\bibinfo {author} {\bibfnamefont {T.}~\bibnamefont
  {P{\"{o}}schel}}, \bibinfo {author} {\bibfnamefont {T.}~\bibnamefont
  {Schwager}},\ and\ \bibinfo {author} {\bibfnamefont {N.~V.}\ \bibnamefont
  {Brilliantov}},\ }\bibfield  {title} {\bibinfo {title} {{Rolling friction of
  a hard cylinder on a viscous plane}},\ }\href
  {https://doi.org/10.1007/s100510050840} {\bibfield  {journal} {\bibinfo
  {journal} {Eur. Phys. J B}\ }\textbf {\bibinfo {volume} {10}},\ \bibinfo
  {pages} {169} (\bibinfo {year} {1999})}\BibitemShut {NoStop}%
\bibitem [{\citenamefont {Johnson}(1985)}]{Johnson1985}%
  \BibitemOpen
  \bibfield  {author} {\bibinfo {author} {\bibfnamefont {K.~L.}\ \bibnamefont
  {Johnson}},\ }\href@noop {} {\emph {\bibinfo {title} {{Contact mechanics}}}}\
  (\bibinfo  {publisher} {Cambridge University Press},\ \bibinfo {address}
  {Cambridge},\ \bibinfo {year} {1985})\BibitemShut {NoStop}%
\bibitem [{\citenamefont {Silbert}\ \emph {et~al.}(2002)\citenamefont
  {Silbert}, \citenamefont {Ertas}, \citenamefont {Grest}, \citenamefont
  {Halsey},\ and\ \citenamefont {Levine}}]{Silbert2002}%
  \BibitemOpen
  \bibfield  {author} {\bibinfo {author} {\bibfnamefont {L.~E.}\ \bibnamefont
  {Silbert}}, \bibinfo {author} {\bibfnamefont {D.}~\bibnamefont {Ertas}},
  \bibinfo {author} {\bibfnamefont {G.~S.}\ \bibnamefont {Grest}}, \bibinfo
  {author} {\bibfnamefont {T.~C.}\ \bibnamefont {Halsey}},\ and\ \bibinfo
  {author} {\bibfnamefont {D.}~\bibnamefont {Levine}},\ }\bibfield  {title}
  {\bibinfo {title} {{Geometry of frictionless and frictional sphere
  packings}},\ }\href {https://doi.org/10.1103/PhysRevE.65.031304} {\bibfield
  {journal} {\bibinfo  {journal} {Phys. Rev. E}\ }\textbf {\bibinfo {volume}
  {65}},\ \bibinfo {pages} {031304} (\bibinfo {year} {2002})}\BibitemShut
  {NoStop}%
\bibitem [{\citenamefont {Shundyak}\ \emph {et~al.}(2007)\citenamefont
  {Shundyak}, \citenamefont {{van Hecke}},\ and\ \citenamefont {{van
  Saarloos}}}]{Shundyak2007}%
  \BibitemOpen
  \bibfield  {author} {\bibinfo {author} {\bibfnamefont {K.}~\bibnamefont
  {Shundyak}}, \bibinfo {author} {\bibfnamefont {M.}~\bibnamefont {{van
  Hecke}}},\ and\ \bibinfo {author} {\bibfnamefont {W.}~\bibnamefont {{van
  Saarloos}}},\ }\bibfield  {title} {\bibinfo {title} {{Force mobilization and
  generalized isostaticity in jammed packings of frictional grains}},\ }\href
  {https://doi.org/10.1103/PhysRevE.75.010301} {\bibfield  {journal} {\bibinfo
  {journal} {Phys. Rev. E}\ }\textbf {\bibinfo {volume} {75}},\ \bibinfo
  {pages} {010301(R)} (\bibinfo {year} {2007})}\BibitemShut {NoStop}%
\bibitem [{\citenamefont {Somfai}\ \emph {et~al.}(2007)\citenamefont {Somfai},
  \citenamefont {{van Hecke}}, \citenamefont {Ellenbroek}, \citenamefont
  {Shundyak},\ and\ \citenamefont {{van Saarloos}}}]{Somfai2007}%
  \BibitemOpen
  \bibfield  {author} {\bibinfo {author} {\bibfnamefont {E.}~\bibnamefont
  {Somfai}}, \bibinfo {author} {\bibfnamefont {M.}~\bibnamefont {{van Hecke}}},
  \bibinfo {author} {\bibfnamefont {W.~G.}\ \bibnamefont {Ellenbroek}},
  \bibinfo {author} {\bibfnamefont {K.}~\bibnamefont {Shundyak}},\ and\
  \bibinfo {author} {\bibfnamefont {W.}~\bibnamefont {{van Saarloos}}},\
  }\bibfield  {title} {\bibinfo {title} {{Critical and noncritical jamming of
  frictional grains}},\ }\href {https://doi.org/10.1103/PhysRevE.75.020301}
  {\bibfield  {journal} {\bibinfo  {journal} {Phys. Rev. E}\ }\textbf {\bibinfo
  {volume} {75}},\ \bibinfo {pages} {020301(R)} (\bibinfo {year}
  {2007})}\BibitemShut {NoStop}%
\bibitem [{\citenamefont {Song}\ \emph {et~al.}(2008)\citenamefont {Song},
  \citenamefont {Wang},\ and\ \citenamefont {Makse}}]{Song2008a}%
  \BibitemOpen
  \bibfield  {author} {\bibinfo {author} {\bibfnamefont {C.}~\bibnamefont
  {Song}}, \bibinfo {author} {\bibfnamefont {P.}~\bibnamefont {Wang}},\ and\
  \bibinfo {author} {\bibfnamefont {H.~A.}\ \bibnamefont {Makse}},\ }\bibfield
  {title} {\bibinfo {title} {{A phase diagram for jammed matter}},\ }\href
  {https://doi.org/10.1038/nature06981} {\bibfield  {journal} {\bibinfo
  {journal} {Nature}\ }\textbf {\bibinfo {volume} {453}},\ \bibinfo {pages}
  {629} (\bibinfo {year} {2008})}\BibitemShut {NoStop}%
\bibitem [{\citenamefont {Silbert}(2010)}]{Silbert2010}%
  \BibitemOpen
  \bibfield  {author} {\bibinfo {author} {\bibfnamefont {L.~E.}\ \bibnamefont
  {Silbert}},\ }\bibfield  {title} {\bibinfo {title} {{Jamming of frictional
  spheres and random loose packing}},\ }\href
  {https://doi.org/10.1039/c001973a} {\bibfield  {journal} {\bibinfo  {journal}
  {Soft Matter}\ }\textbf {\bibinfo {volume} {6}},\ \bibinfo {pages} {2918}
  (\bibinfo {year} {2010})}\BibitemShut {NoStop}%
\bibitem [{\citenamefont {Torquato}\ \emph {et~al.}(2000)\citenamefont
  {Torquato}, \citenamefont {Truskett},\ and\ \citenamefont
  {Debenedetti}}]{Torquato2000}%
  \BibitemOpen
  \bibfield  {author} {\bibinfo {author} {\bibfnamefont {S.}~\bibnamefont
  {Torquato}}, \bibinfo {author} {\bibfnamefont {T.~M.}\ \bibnamefont
  {Truskett}},\ and\ \bibinfo {author} {\bibfnamefont {P.~G.}\ \bibnamefont
  {Debenedetti}},\ }\bibfield  {title} {\bibinfo {title} {{Is Random Close
  Packing of Spheres Well Defined?}},\ }\href@noop {} {\bibfield  {journal}
  {\bibinfo  {journal} {Phys. Rev. Lett.}\ }\textbf {\bibinfo {volume} {84}},\
  \bibinfo {pages} {2064} (\bibinfo {year} {2000})}\BibitemShut {NoStop}%
\bibitem [{\citenamefont {Scott}\ and\ \citenamefont
  {Kilgour}(1969)}]{Scott1969}%
  \BibitemOpen
  \bibfield  {author} {\bibinfo {author} {\bibfnamefont {G.~D.}\ \bibnamefont
  {Scott}}\ and\ \bibinfo {author} {\bibfnamefont {D.~M.}\ \bibnamefont
  {Kilgour}},\ }\bibfield  {title} {\bibinfo {title} {{The density of random
  close packing of spheres}},\ }\href
  {https://doi.org/10.1088/0022-3727/2/6/311} {\bibfield  {journal} {\bibinfo
  {journal} {J. Phy. D Appl. Phys.}\ }\textbf {\bibinfo {volume} {2}},\
  \bibinfo {pages} {863} (\bibinfo {year} {1969})}\BibitemShut {NoStop}%
\bibitem [{\citenamefont {R.~L.~Brown}\ and\ \citenamefont
  {Hawksley}(1946)}]{Brown1946}%
  \BibitemOpen
  \bibfield  {author} {\bibinfo {author} {\bibfnamefont {R.~L.}\ \bibnamefont
  {R.~L.~Brown}}\ and\ \bibinfo {author} {\bibfnamefont {P.~G.~W.}\
  \bibnamefont {Hawksley}},\ }\bibfield  {title} {\bibinfo {title} {{Effect of
  Container Walls on Packing Density of Particles}},\ }\href
  {https://doi.org/10.1017/CBO9781107415324.004} {\bibfield  {journal}
  {\bibinfo  {journal} {Nature}\ }\textbf {\bibinfo {volume} {157}},\ \bibinfo
  {pages} {585} (\bibinfo {year} {1946})}\BibitemShut {NoStop}%
\bibitem [{\citenamefont {Scott}(1960)}]{Scott1960}%
  \BibitemOpen
  \bibfield  {author} {\bibinfo {author} {\bibfnamefont {G.~D.}\ \bibnamefont
  {Scott}},\ }\bibfield  {title} {\bibinfo {title} {{Packing of spheres:
  Packing of equal spheres}},\ }\href {https://doi.org/10.1038/188908a0}
  {\bibfield  {journal} {\bibinfo  {journal} {Nature}\ }\textbf {\bibinfo
  {volume} {188}},\ \bibinfo {pages} {908} (\bibinfo {year}
  {1960})}\BibitemShut {NoStop}%
\bibitem [{\citenamefont {Rutgers}(1962)}]{Rutgers1962}%
  \BibitemOpen
  \bibfield  {author} {\bibinfo {author} {\bibfnamefont {R.}~\bibnamefont
  {Rutgers}},\ }\bibfield  {title} {\bibinfo {title} {{Packing of sphere}},\
  }\href {https://doi.org/10.1038/193465a0} {\bibfield  {journal} {\bibinfo
  {journal} {Nature}\ }\textbf {\bibinfo {volume} {193}},\ \bibinfo {pages}
  {465} (\bibinfo {year} {1962})}\BibitemShut {NoStop}%
\bibitem [{\citenamefont {Jerkins}\ \emph {et~al.}(2008)\citenamefont
  {Jerkins}, \citenamefont {Schroter}, \citenamefont {Swinney}, \citenamefont
  {Senden}, \citenamefont {Saadatfar},\ and\ \citenamefont
  {Aste}}]{Jerkins2008}%
  \BibitemOpen
  \bibfield  {author} {\bibinfo {author} {\bibfnamefont {M.}~\bibnamefont
  {Jerkins}}, \bibinfo {author} {\bibfnamefont {M.}~\bibnamefont {Schroter}},
  \bibinfo {author} {\bibfnamefont {H.~L.}\ \bibnamefont {Swinney}}, \bibinfo
  {author} {\bibfnamefont {T.~J.}\ \bibnamefont {Senden}}, \bibinfo {author}
  {\bibfnamefont {M.}~\bibnamefont {Saadatfar}},\ and\ \bibinfo {author}
  {\bibfnamefont {T.}~\bibnamefont {Aste}},\ }\bibfield  {title} {\bibinfo
  {title} {{Onset of mechanical stability in random packings of frictional
  spheres}},\ }\href {https://doi.org/10.1103/PhysRevLett.101.018301}
  {\bibfield  {journal} {\bibinfo  {journal} {Phys. Rev. Lett.}\ }\textbf
  {\bibinfo {volume} {101}},\ \bibinfo {pages} {018301} (\bibinfo {year}
  {2008})}\BibitemShut {NoStop}%
\bibitem [{\citenamefont {Farrell}\ \emph {et~al.}(2010)\citenamefont
  {Farrell}, \citenamefont {Martini},\ and\ \citenamefont
  {Menon}}]{Farrell2010}%
  \BibitemOpen
  \bibfield  {author} {\bibinfo {author} {\bibfnamefont {G.~R.}\ \bibnamefont
  {Farrell}}, \bibinfo {author} {\bibfnamefont {K.~M.}\ \bibnamefont
  {Martini}},\ and\ \bibinfo {author} {\bibfnamefont {N.}~\bibnamefont
  {Menon}},\ }\bibfield  {title} {\bibinfo {title} {{Loose packings of
  frictional spheres}},\ }\href {https://doi.org/10.1039/c0sm00038h} {\bibfield
   {journal} {\bibinfo  {journal} {Soft Matter}\ }\textbf {\bibinfo {volume}
  {6}},\ \bibinfo {pages} {2925} (\bibinfo {year} {2010})}\BibitemShut
  {NoStop}%
\bibitem [{\citenamefont {Dagois-Bohy}\ \emph {et~al.}(2012)\citenamefont
  {Dagois-Bohy}, \citenamefont {Tighe}, \citenamefont {Simon}, \citenamefont
  {Henkes},\ and\ \citenamefont {{van Hecke}}}]{Dagois-Bohy2012}%
  \BibitemOpen
  \bibfield  {author} {\bibinfo {author} {\bibfnamefont {S.}~\bibnamefont
  {Dagois-Bohy}}, \bibinfo {author} {\bibfnamefont {B.~P.}\ \bibnamefont
  {Tighe}}, \bibinfo {author} {\bibfnamefont {J.}~\bibnamefont {Simon}},
  \bibinfo {author} {\bibfnamefont {S.}~\bibnamefont {Henkes}},\ and\ \bibinfo
  {author} {\bibfnamefont {M.}~\bibnamefont {{van Hecke}}},\ }\bibfield
  {title} {\bibinfo {title} {{Soft-sphere packings at finite pressure but
  unstable to shear}},\ }\href {https://doi.org/10.1103/PhysRevLett.109.095703}
  {\bibfield  {journal} {\bibinfo  {journal} {Phys. Rev. Lett.}\ }\textbf
  {\bibinfo {volume} {109}},\ \bibinfo {pages} {095703} (\bibinfo {year}
  {2012})}\BibitemShut {NoStop}%
\bibitem [{\citenamefont {Smith}\ \emph {et~al.}(2014)\citenamefont {Smith},
  \citenamefont {Srivastava}, \citenamefont {Fisher},\ and\ \citenamefont
  {Alam}}]{Smith2014}%
  \BibitemOpen
  \bibfield  {author} {\bibinfo {author} {\bibfnamefont {K.~C.}\ \bibnamefont
  {Smith}}, \bibinfo {author} {\bibfnamefont {I.}~\bibnamefont {Srivastava}},
  \bibinfo {author} {\bibfnamefont {T.~S.}\ \bibnamefont {Fisher}},\ and\
  \bibinfo {author} {\bibfnamefont {M.}~\bibnamefont {Alam}},\ }\bibfield
  {title} {\bibinfo {title} {{Variable-cell method for stress-controlled
  jamming of athermal, frictionless grains}},\ }\href
  {https://doi.org/10.1103/PhysRevE.89.042203} {\bibfield  {journal} {\bibinfo
  {journal} {Phys. Rev. E}\ }\textbf {\bibinfo {volume} {89}},\ \bibinfo
  {pages} {042203} (\bibinfo {year} {2014})}\BibitemShut {NoStop}%
\bibitem [{\citenamefont {Srivastava}\ and\ \citenamefont
  {Fisher}(2017)}]{Srivastava2017}%
  \BibitemOpen
  \bibfield  {author} {\bibinfo {author} {\bibfnamefont {I.}~\bibnamefont
  {Srivastava}}\ and\ \bibinfo {author} {\bibfnamefont {T.~S.}\ \bibnamefont
  {Fisher}},\ }\bibfield  {title} {\bibinfo {title} {{Slow creep in soft
  granular packings}},\ }\href {https://doi.org/10.1039/c7sm00237h} {\bibfield
  {journal} {\bibinfo  {journal} {Soft Matter}\ }\textbf {\bibinfo {volume}
  {13}},\ \bibinfo {pages} {3411} (\bibinfo {year} {2017})}\BibitemShut
  {NoStop}%
\bibitem [{\citenamefont {Cundall}\ and\ \citenamefont
  {Strack}(1979)}]{Cundall1979}%
  \BibitemOpen
  \bibfield  {author} {\bibinfo {author} {\bibfnamefont {P.~A.}\ \bibnamefont
  {Cundall}}\ and\ \bibinfo {author} {\bibfnamefont {O.~D.~L.}\ \bibnamefont
  {Strack}},\ }\bibfield  {title} {\bibinfo {title} {{A discrete numerical
  model for granular assemblies}},\ }\href@noop {} {\bibfield  {journal}
  {\bibinfo  {journal} {Geotechnique}\ }\textbf {\bibinfo {volume} {29}},\
  \bibinfo {pages} {47} (\bibinfo {year} {1979})}\BibitemShut {NoStop}%
\bibitem [{\citenamefont {Luding}(2008)}]{Luding2008}%
  \BibitemOpen
  \bibfield  {author} {\bibinfo {author} {\bibfnamefont {S.}~\bibnamefont
  {Luding}},\ }\bibfield  {title} {\bibinfo {title} {{Cohesive, frictional
  powders: Contact models for tension}},\ }\href
  {https://doi.org/10.1007/s10035-008-0099-x} {\bibfield  {journal} {\bibinfo
  {journal} {Granular Matter}\ }\textbf {\bibinfo {volume} {10}},\ \bibinfo
  {pages} {235} (\bibinfo {year} {2008})}\BibitemShut {NoStop}%
\bibitem [{\citenamefont {Marshall}(2009)}]{Marshall2009}%
  \BibitemOpen
  \bibfield  {author} {\bibinfo {author} {\bibfnamefont {J.~S.}\ \bibnamefont
  {Marshall}},\ }\bibfield  {title} {\bibinfo {title} {{Discrete-element
  modeling of particulate aerosol flows}},\ }\href
  {https://doi.org/10.1016/j.jcp.2008.10.035} {\bibfield  {journal} {\bibinfo
  {journal} {J. Comput.Phys.}\ }\textbf {\bibinfo {volume} {228}},\ \bibinfo
  {pages} {1541} (\bibinfo {year} {2009})}\BibitemShut {NoStop}%
\bibitem [{\citenamefont {Foerster}\ \emph {et~al.}(1994)\citenamefont
  {Foerster}, \citenamefont {Louge}, \citenamefont {Chang},\ and\ \citenamefont
  {Allia}}]{Foerster1994}%
  \BibitemOpen
  \bibfield  {author} {\bibinfo {author} {\bibfnamefont {S.~F.}\ \bibnamefont
  {Foerster}}, \bibinfo {author} {\bibfnamefont {M.~Y.}\ \bibnamefont {Louge}},
  \bibinfo {author} {\bibfnamefont {H.}~\bibnamefont {Chang}},\ and\ \bibinfo
  {author} {\bibfnamefont {K.}~\bibnamefont {Allia}},\ }\bibfield  {title}
  {\bibinfo {title} {{Measurements of the collision properties of small
  spheres}},\ }\href@noop {} {\bibfield  {journal} {\bibinfo  {journal} {Phys.
  Fluids}\ }\textbf {\bibinfo {volume} {6}},\ \bibinfo {pages} {1108} (\bibinfo
  {year} {1994})}\BibitemShut {NoStop}%
\bibitem [{Note1()}]{Note1}%
  \BibitemOpen
  \bibinfo {note} {To use this interaction model in LAMMPS~\cite
  {StevePlimton1995}, use the following commands: \protect \texttt
  {pair\protect \_style granular} followed by \protect \texttt {pair\protect
  \_coeff * * hooke 1 0.5 damping mass\protect \_velocity tangential
  linear\protect \_history 1 1 $\mu _s$ rolling sds 1 0.5 $\mu _r$ twisting sds
  1 0.5 $\mu _t$}. See LAMMPS documentation at lammps.sandia.gov for more
  details.}\BibitemShut {Stop}%
\bibitem [{\citenamefont {Tabor}(1955)}]{Tabor1955}%
  \BibitemOpen
  \bibfield  {author} {\bibinfo {author} {\bibfnamefont {D.}~\bibnamefont
  {Tabor}},\ }\bibfield  {title} {\bibinfo {title} {{The mechanism of rolling
  friction. II. The elastic range}},\ }\href@noop {} {\bibfield  {journal}
  {\bibinfo  {journal} {Proc. R. Soc. Lond. A}\ }\textbf {\bibinfo {volume}
  {229}},\ \bibinfo {pages} {198} (\bibinfo {year} {1955})}\BibitemShut
  {NoStop}%
\bibitem [{\citenamefont {Halling}(1959)}]{Halling1959}%
  \BibitemOpen
  \bibfield  {author} {\bibinfo {author} {\bibfnamefont {J.}~\bibnamefont
  {Halling}},\ }\bibfield  {title} {\bibinfo {title} {{Effect of deformation of
  the surface texture on rolling resistance}},\ }\href
  {https://doi.org/10.1088/0508-3443/10/4/304} {\bibfield  {journal} {\bibinfo
  {journal} {British J. Appl. Phys.}\ }\textbf {\bibinfo {volume} {10}},\
  \bibinfo {pages} {172} (\bibinfo {year} {1959})}\BibitemShut {NoStop}%
\bibitem [{\citenamefont {Carbone}\ and\ \citenamefont
  {Putignano}(2013)}]{Carbone2013}%
  \BibitemOpen
  \bibfield  {author} {\bibinfo {author} {\bibfnamefont {G.}~\bibnamefont
  {Carbone}}\ and\ \bibinfo {author} {\bibfnamefont {C.}~\bibnamefont
  {Putignano}},\ }\bibfield  {title} {\bibinfo {title} {{A novel methodology to
  predict sliding and rolling friction of viscoelastic materials: Theory and
  experiments}},\ }\href {https://doi.org/10.1016/j.jmps.2013.03.005}
  {\bibfield  {journal} {\bibinfo  {journal} {J. Mech. Phys. Solids}\ }\textbf
  {\bibinfo {volume} {61}},\ \bibinfo {pages} {1822} (\bibinfo {year}
  {2013})}\BibitemShut {NoStop}%
\bibitem [{\citenamefont {Plimpton}(1995)}]{StevePlimton1995}%
  \BibitemOpen
  \bibfield  {author} {\bibinfo {author} {\bibfnamefont {S.}~\bibnamefont
  {Plimpton}},\ }\bibfield  {title} {\bibinfo {title} {{Fast Parallel
  Algorithms for Short-Range Molecular Dynamics}},\ }\href@noop {} {\bibfield
  {journal} {\bibinfo  {journal} {J. Comput.Phys.}\ }\textbf {\bibinfo {volume}
  {117}},\ \bibinfo {pages} {1} (\bibinfo {year} {1995})}\BibitemShut {NoStop}%
\bibitem [{\citenamefont {Shinoda}\ \emph {et~al.}(2004)\citenamefont
  {Shinoda}, \citenamefont {Shiga},\ and\ \citenamefont
  {Mikami}}]{Shinoda2004}%
  \BibitemOpen
  \bibfield  {author} {\bibinfo {author} {\bibfnamefont {W.}~\bibnamefont
  {Shinoda}}, \bibinfo {author} {\bibfnamefont {M.}~\bibnamefont {Shiga}},\
  and\ \bibinfo {author} {\bibfnamefont {M.}~\bibnamefont {Mikami}},\
  }\bibfield  {title} {\bibinfo {title} {{Rapid estimation of elastic constants
  by molecular dynamics simulation under constant stress}},\ }\href
  {https://doi.org/10.1103/PhysRevB.69.134103} {\bibfield  {journal} {\bibinfo
  {journal} {Phys. Rev. B}\ }\textbf {\bibinfo {volume} {69}},\ \bibinfo
  {pages} {134103} (\bibinfo {year} {2004})}\BibitemShut {NoStop}%
\bibitem [{\citenamefont {Martyna}\ \emph {et~al.}(1994)\citenamefont
  {Martyna}, \citenamefont {Tobias},\ and\ \citenamefont
  {Klein}}]{Martyna1994}%
  \BibitemOpen
  \bibfield  {author} {\bibinfo {author} {\bibfnamefont {G.~J.}\ \bibnamefont
  {Martyna}}, \bibinfo {author} {\bibfnamefont {D.~J.}\ \bibnamefont
  {Tobias}},\ and\ \bibinfo {author} {\bibfnamefont {M.~L.}\ \bibnamefont
  {Klein}},\ }\bibfield  {title} {\bibinfo {title} {{Constant pressure
  molecular dynamics algorithms}},\ }\href@noop {} {\bibfield  {journal}
  {\bibinfo  {journal} {J. Chem. Phys.}\ }\textbf {\bibinfo {volume} {101}},\
  \bibinfo {pages} {4177} (\bibinfo {year} {1994})}\BibitemShut {NoStop}%
\bibitem [{\citenamefont {Parrinello}\ and\ \citenamefont
  {Rahman}(1981)}]{Parrinello1981}%
  \BibitemOpen
  \bibfield  {author} {\bibinfo {author} {\bibfnamefont {M.}~\bibnamefont
  {Parrinello}}\ and\ \bibinfo {author} {\bibfnamefont {A.}~\bibnamefont
  {Rahman}},\ }\bibfield  {title} {\bibinfo {title} {{Polymorphic transitions
  in single crystals: A new molecular dynamics method}},\ }\href
  {https://doi.org/10.1063/1.328693} {\bibfield  {journal} {\bibinfo  {journal}
  {J. Appl. Phys.}\ }\textbf {\bibinfo {volume} {52}},\ \bibinfo {pages} {7182}
  (\bibinfo {year} {1981})}\BibitemShut {NoStop}%
\bibitem [{Note2()}]{Note2}%
  \BibitemOpen
  \bibinfo {note} {To apply this symmetric pressure tensors in LAMMPS~\cite
  {StevePlimton1995}, use \protect \texttt {fix 1 all nph/sphere x 1e-4 1e-4
  2.25 y 1e-4 1e-4 2.25 z 1e-4 1e-4 2.25 xy 0.0 0.0 2.25 yz 0.0 0.0 2.25 nreset
  1 pchain 0}. See LAMMPS documentation at lammps.sandia.gov for more
  details.}\BibitemShut {Stop}%
\bibitem [{\citenamefont {Maxwell}(1865)}]{Maxwell1865}%
  \BibitemOpen
  \bibfield  {author} {\bibinfo {author} {\bibfnamefont {J.}~\bibnamefont
  {Maxwell}},\ }\bibfield  {title} {\bibinfo {title} {{On the calculation of
  the equilibrium and stiffness of frames}},\ }\href@noop {} {\bibfield
  {journal} {\bibinfo  {journal} {Philos. Mag.}\ }\textbf {\bibinfo {volume}
  {27}},\ \bibinfo {pages} {294} (\bibinfo {year} {1865})}\BibitemShut
  {NoStop}%
\bibitem [{\citenamefont {Liu}\ \emph {et~al.}(2017)\citenamefont {Liu},
  \citenamefont {Jin}, \citenamefont {Chen}, \citenamefont {Makse},\ and\
  \citenamefont {Li}}]{Liu2017b}%
  \BibitemOpen
  \bibfield  {author} {\bibinfo {author} {\bibfnamefont {W.}~\bibnamefont
  {Liu}}, \bibinfo {author} {\bibfnamefont {Y.}~\bibnamefont {Jin}}, \bibinfo
  {author} {\bibfnamefont {S.}~\bibnamefont {Chen}}, \bibinfo {author}
  {\bibfnamefont {H.~A.}\ \bibnamefont {Makse}},\ and\ \bibinfo {author}
  {\bibfnamefont {S.}~\bibnamefont {Li}},\ }\bibfield  {title} {\bibinfo
  {title} {{Equation of state for random sphere packings with arbitrary
  adhesion and friction}},\ }\href {https://doi.org/10.1039/c6sm02216b}
  {\bibfield  {journal} {\bibinfo  {journal} {Soft Matter}\ }\textbf {\bibinfo
  {volume} {13}},\ \bibinfo {pages} {421} (\bibinfo {year} {2017})}\BibitemShut
  {NoStop}%
\bibitem [{\citenamefont {Khan}\ and\ \citenamefont
  {Bushell}(2005)}]{Khan2005}%
  \BibitemOpen
  \bibfield  {author} {\bibinfo {author} {\bibfnamefont {K.~M.}\ \bibnamefont
  {Khan}}\ and\ \bibinfo {author} {\bibfnamefont {G.}~\bibnamefont {Bushell}},\
  }\bibfield  {title} {\bibinfo {title} {{Comment on "rolling friction in the
  dynamic simulation of sandpile formation"}},\ }\href
  {https://doi.org/10.1016/j.physa.2005.01.019} {\bibfield  {journal} {\bibinfo
   {journal} {Physica A}\ }\textbf {\bibinfo {volume} {352}},\ \bibinfo {pages}
  {522} (\bibinfo {year} {2005})}\BibitemShut {NoStop}%
\bibitem [{\citenamefont {Jiang}\ \emph {et~al.}(2006)\citenamefont {Jiang},
  \citenamefont {Yu},\ and\ \citenamefont {Harris}}]{Jiang2006}%
  \BibitemOpen
  \bibfield  {author} {\bibinfo {author} {\bibfnamefont {M.~J.}\ \bibnamefont
  {Jiang}}, \bibinfo {author} {\bibfnamefont {H.~S.}\ \bibnamefont {Yu}},\ and\
  \bibinfo {author} {\bibfnamefont {D.}~\bibnamefont {Harris}},\ }\bibfield
  {title} {\bibinfo {title} {{Bond rolling resistance and its effect on
  yielding of bonded granulates by DEM analyses}},\ }\href
  {https://doi.org/10.1002/nag.498} {\bibfield  {journal} {\bibinfo  {journal}
  {Int. J. Numer. Anal. Meth. Geomech.}\ }\textbf {\bibinfo {volume} {30}},\
  \bibinfo {pages} {723} (\bibinfo {year} {2006})}\BibitemShut {NoStop}%
\bibitem [{\citenamefont {Zhu}\ and\ \citenamefont {Yu}(2006)}]{Zhu2006}%
  \BibitemOpen
  \bibfield  {author} {\bibinfo {author} {\bibfnamefont {H.~P.}\ \bibnamefont
  {Zhu}}\ and\ \bibinfo {author} {\bibfnamefont {A.~B.}\ \bibnamefont {Yu}},\
  }\bibfield  {title} {\bibinfo {title} {{A theoretical analysis of the force
  models in discrete element method}},\ }\href
  {https://doi.org/10.1016/j.powtec.2005.09.006} {\bibfield  {journal}
  {\bibinfo  {journal} {Powder Technol.}\ }\textbf {\bibinfo {volume} {161}},\
  \bibinfo {pages} {122} (\bibinfo {year} {2006})}\BibitemShut {NoStop}%
\bibitem [{\citenamefont {Ai}\ \emph {et~al.}(2011)\citenamefont {Ai},
  \citenamefont {Chen}, \citenamefont {Rotter},\ and\ \citenamefont
  {Ooi}}]{Ai2011}%
  \BibitemOpen
  \bibfield  {author} {\bibinfo {author} {\bibfnamefont {J.}~\bibnamefont
  {Ai}}, \bibinfo {author} {\bibfnamefont {J.~F.}\ \bibnamefont {Chen}},
  \bibinfo {author} {\bibfnamefont {J.~M.}\ \bibnamefont {Rotter}},\ and\
  \bibinfo {author} {\bibfnamefont {J.~Y.}\ \bibnamefont {Ooi}},\ }\bibfield
  {title} {\bibinfo {title} {{Assessment of rolling resistance models in
  discrete element simulations}},\ }\href
  {https://doi.org/10.1016/j.powtec.2010.09.030} {\bibfield  {journal}
  {\bibinfo  {journal} {Powder Technol.}\ }\textbf {\bibinfo {volume} {206}},\
  \bibinfo {pages} {269} (\bibinfo {year} {2011})}\BibitemShut {NoStop}%
\bibitem [{\citenamefont {Wang}\ \emph {et~al.}(2015)\citenamefont {Wang},
  \citenamefont {Alonso-Marroquin},\ and\ \citenamefont {Guo}}]{Wang2015c}%
  \BibitemOpen
  \bibfield  {author} {\bibinfo {author} {\bibfnamefont {Y.}~\bibnamefont
  {Wang}}, \bibinfo {author} {\bibfnamefont {F.}~\bibnamefont
  {Alonso-Marroquin}},\ and\ \bibinfo {author} {\bibfnamefont {W.~W.}\
  \bibnamefont {Guo}},\ }\bibfield  {title} {\bibinfo {title} {{Rolling and
  sliding in 3-D discrete element models}},\ }\href
  {https://doi.org/10.1016/j.partic.2015.01.006} {\bibfield  {journal}
  {\bibinfo  {journal} {Particuology}\ }\textbf {\bibinfo {volume} {23}},\
  \bibinfo {pages} {49} (\bibinfo {year} {2015})}\BibitemShut {NoStop}%
\bibitem [{\citenamefont {Eldredge}\ and\ \citenamefont
  {Tabor}(1955)}]{Eldredge1955}%
  \BibitemOpen
  \bibfield  {author} {\bibinfo {author} {\bibfnamefont {K.~R.}\ \bibnamefont
  {Eldredge}}\ and\ \bibinfo {author} {\bibfnamefont {D.}~\bibnamefont
  {Tabor}},\ }\bibfield  {title} {\bibinfo {title} {{The mechanism of rolling
  friction. I. The plastic range}},\ }\href@noop {} {\bibfield  {journal}
  {\bibinfo  {journal} {Proc. R. Soc. Lond. A}\ }\textbf {\bibinfo {volume}
  {229}},\ \bibinfo {pages} {181} (\bibinfo {year} {1955})}\BibitemShut
  {NoStop}%
\bibitem [{\citenamefont {Drutowski}(1959)}]{Drutowski1959}%
  \BibitemOpen
  \bibfield  {author} {\bibinfo {author} {\bibfnamefont {R.~C.}\ \bibnamefont
  {Drutowski}},\ }\bibfield  {title} {\bibinfo {title} {{Energy losses of balls
  rolling on plates}},\ }\href@noop {} {\bibfield  {journal} {\bibinfo
  {journal} {J. Basic Eng.}\ }\textbf {\bibinfo {volume} {81}},\ \bibinfo
  {pages} {233} (\bibinfo {year} {1959})}\BibitemShut {NoStop}%
\bibitem [{\citenamefont {May}\ \emph {et~al.}(1959)\citenamefont {May},
  \citenamefont {Morris},\ and\ \citenamefont {Atack}}]{May1959}%
  \BibitemOpen
  \bibfield  {author} {\bibinfo {author} {\bibfnamefont {W.~D.}\ \bibnamefont
  {May}}, \bibinfo {author} {\bibfnamefont {E.~L.}\ \bibnamefont {Morris}},\
  and\ \bibinfo {author} {\bibfnamefont {D.}~\bibnamefont {Atack}},\ }\bibfield
   {title} {\bibinfo {title} {{Rolling friction of a hard cylinder over a
  viscoelastic material}},\ }\href {https://doi.org/10.1063/1.1735042}
  {\bibfield  {journal} {\bibinfo  {journal} {J. Appl. Phys.}\ }\textbf
  {\bibinfo {volume} {30}},\ \bibinfo {pages} {1713} (\bibinfo {year}
  {1959})}\BibitemShut {NoStop}%
\bibitem [{\citenamefont {Flom}\ and\ \citenamefont {Bueche}(1959)}]{Flom1959}%
  \BibitemOpen
  \bibfield  {author} {\bibinfo {author} {\bibfnamefont {D.~G.}\ \bibnamefont
  {Flom}}\ and\ \bibinfo {author} {\bibfnamefont {A.~M.}\ \bibnamefont
  {Bueche}},\ }\bibfield  {title} {\bibinfo {title} {{Theory of rolling
  friction for spheres}},\ }\href {https://doi.org/10.1063/1.1735043}
  {\bibfield  {journal} {\bibinfo  {journal} {J. Appl. Phys.}\ }\textbf
  {\bibinfo {volume} {30}},\ \bibinfo {pages} {1725} (\bibinfo {year}
  {1959})}\BibitemShut {NoStop}%
\bibitem [{\citenamefont {Hunter}(1960)}]{Hunter1960}%
  \BibitemOpen
  \bibfield  {author} {\bibinfo {author} {\bibfnamefont {S.~C.}\ \bibnamefont
  {Hunter}},\ }\bibfield  {title} {\bibinfo {title} {{The Hertz problem for a
  rigid spherical indenter and a viscoelastic half-space}},\ }\href
  {https://doi.org/10.1016/0022-5096(60)90028-4} {\bibfield  {journal}
  {\bibinfo  {journal} {J. Mech. Phys. Solids}\ }\textbf {\bibinfo {volume}
  {8}},\ \bibinfo {pages} {219} (\bibinfo {year} {1960})}\BibitemShut {NoStop}%
\bibitem [{\citenamefont {Jiang}\ \emph {et~al.}(2005)\citenamefont {Jiang},
  \citenamefont {Yu},\ and\ \citenamefont {Harris}}]{Jiang2005}%
  \BibitemOpen
  \bibfield  {author} {\bibinfo {author} {\bibfnamefont {M.~J.}\ \bibnamefont
  {Jiang}}, \bibinfo {author} {\bibfnamefont {H.~S.}\ \bibnamefont {Yu}},\ and\
  \bibinfo {author} {\bibfnamefont {D.}~\bibnamefont {Harris}},\ }\bibfield
  {title} {\bibinfo {title} {{A novel discrete model for granular material
  incorporating rolling resistance}},\ }\href
  {https://doi.org/10.1016/j.compgeo.2005.05.001} {\bibfield  {journal}
  {\bibinfo  {journal} {Comput. Geol.}\ }\textbf {\bibinfo {volume} {32}},\
  \bibinfo {pages} {340} (\bibinfo {year} {2005})}\BibitemShut {NoStop}%
\bibitem [{\citenamefont {Lubkin}(1951)}]{Lubkin1951}%
  \BibitemOpen
  \bibfield  {author} {\bibinfo {author} {\bibfnamefont {J.~L.}\ \bibnamefont
  {Lubkin}},\ }\bibfield  {title} {\bibinfo {title} {{The torsion of elastic
  spheres in contact}},\ }\href@noop {} {\bibfield  {journal} {\bibinfo
  {journal} {ASME Trans. J. App. Mech.}\ }\textbf {\bibinfo {volume} {18}},\
  \bibinfo {pages} {183} (\bibinfo {year} {1951})}\BibitemShut {NoStop}%
\bibitem [{\citenamefont {Johnson}(1958{\natexlab{a}})}]{Johnson1958a}%
  \BibitemOpen
  \bibfield  {author} {\bibinfo {author} {\bibfnamefont {K.~L.}\ \bibnamefont
  {Johnson}},\ }\bibfield  {title} {\bibinfo {title} {{The effect of spin upon
  the rolling motion of an elastic sphere on a plane}},\ }\href@noop {}
  {\bibfield  {journal} {\bibinfo  {journal} {J. Appl. Mech.}\ }\textbf
  {\bibinfo {volume} {25}},\ \bibinfo {pages} {258} (\bibinfo {year}
  {1958}{\natexlab{a}})}\BibitemShut {NoStop}%
\bibitem [{\citenamefont {Johnson}(1958{\natexlab{b}})}]{Johnson1958}%
  \BibitemOpen
  \bibfield  {author} {\bibinfo {author} {\bibfnamefont {K.~L.}\ \bibnamefont
  {Johnson}},\ }\bibfield  {title} {\bibinfo {title} {{The effect of a
  tangential contact force upon the rolling motion of an elastic sphere on a
  plane}},\ }\href@noop {} {\bibfield  {journal} {\bibinfo  {journal} {J. Appl.
  Mech.}\ }\textbf {\bibinfo {volume} {25}},\ \bibinfo {pages} {260} (\bibinfo
  {year} {1958}{\natexlab{b}})}\BibitemShut {NoStop}%
\end{thebibliography}%
\end{document}